# Interplay between spin wave and magnetic vortex

Zhongchen Gao[1,*], Feifei Wang[1], Xiangyong Zhao[1], Tao Wang[1], Jingguo Hu[2], and Peng Yan[3]

[1]Mathematics and Science College, Shanghai Normal University, Shanghai 200233, China
[2]School of Physical Science and Technology, Yangzhou University, Yangzhou 225002, China
[3]School of Electronic Science and Engineering and State Key Laboratory of Electronic Thin Films and Integrated Devices, University of Electronic Science and Technology of China, Chengdu 610054, China

[*]Corresponding author: gaozc0129@shnu.edu.cn;

In this paper, the interplay between spin wave and magnetic vortex is studied. We find three types of magnon scatterings: skew scattering, symmetric side deflection and back reflection, which associate with respectively magnetic topology, energy density distribution and linear momentum transfer torque within vortex. The vortex core exhibits two translational modes: the intrinsic circular mode and a coercive elliptical mode, which can be excited based on permanent and periodic magnon spin-transfer torque effects of spin wave. Lastly, we propose a vortex-based spin wave valve in which via inhomogeneity modulation we access arbitrary control of the phase shift.

## I. Introduction

Spin waves or magnons are fundamental collective excitations of the spin order [1,2,3] and have been considered as ideal information carriers taking advantages of their free of charge transport (hence low energy dissipation) and orders-of-magnitude smaller wavelength compared with electromagnetic waves (hence downscale devices) [4,5]. The excitation, propagation, control and detection of spin waves have opened up the field of magnonics [4-6]. On the other hand, vortex represents the simplest pattern of domain configuration in a uniformly magnetized particle with in-plane curling magnetization around an out-of-plane magnetization at the core [7,8]. Such structure is topologically protected (commonly referred to as magnetic soliton [9]) and characterized by the binary properties of chirality **C** and polarity **P**, each of which suggests an independent bit of information in future high-density memory storage and logic devices [10,11]. Study of the interplay between spin wave and vortex is interesting in two aspects: i). spin wave and vortex structure together make up a highly nonlinear ferromagnetic system which provides an excellent platform to explore the rich physics in nonlinear science; ii). it holds intimate connection to technological development of both magnonic devices and vortex-based applications (notably, vortex-based spin wave excitation [12-15], transportation [16-18] and manipulation [19-22] as well as spin wave-involved vortex stability [23], reversal [24-28] and dynamics [29-31] have been given considerable attention) and finally facilitates the realization of purely magnetic computing scheme [32,33].

Early studies on the magnon-vortex interaction mainly focused on spin waves that belong to the eigenmodes of the magnetic vortex. Specifically, there are radical spin wave mode (with diametric nodes *n*) typically driven by out-of-plane field and azimuthal spin wave mode (with circular nodes *m*) typically excited by in-plane field [34-41]. Relevant findings partly include: Park *et al*. studied the interaction of azimuthal mode with vortex, and found that the azimuthal mode can split into a doublet [22,30]. Such splitting later was found to be a result of dynamic hybridization of two regular azimuthal dipolar modes and a Goldstone-like gyrotropic mode [21]. By taking advantage of such splitting, Kammerer *et al*. reported the polarization-selective vortex core reversal with excitation of the azimuthal modes [25]. On the other hand, Choi *et al*. demonstrated strong radiation of radical spin wave through frequent reversal of vortex core [14,42]. Wintz *et al*. observed radical spin waves with linear and non-reciprocal dispersion based on the gyrations in magnetic vortex pairs [12]. Yoo *et al*. further found that the radical spin wave eigenmodes can assist the vortex core switching [24]. In some other researches, Buess *et al*. obtained simultaneous excitations of radical and azimuthal spin waves based on fast perturbation to overall vortex magnetization [43]. Bauer *et al*. revealed the interference effect between radical mode and azimuthal mode and a variety of nonlinear effects between the spin waves and an initially gyrating core [44]. Ruiz *et al*. discovered the curling nature of vortex states upon concurrent excitations of gyrotropic core dynamics, spiral spin waves, azimuthal and radial modes [45]. Most recently, Wang *et al*. reported that the three-magnon confluence and splitting scattering of azimuthal spin waves off the gyrating vortex core can generate a set of discrete and equally spaced spectral lines, hence a twisted magnon frequency comb [46].

Despite the fruitful results achieved, the demand of long-distance travelling spin wave for information transmission has been largely overlooked. Note that the radical and azimuthal spin waves are commonly locally excited and confined within isolated geometry (e.g., disk or dot). However, in magnonics, following the pace of CMOS technology, practical applications such as logic gates and memories need to be cascaded and form circuitry. Interconnects are key elements for any circuit. In the case of magnonic circuit [47], these interconnects are often referred as spin wave bus, which can be manufactured based on magnetic thin-film nanostrips and undertake to input, transport and output magnonic signals. Naturally, the mutual effect between the spin-wave current and subsequent magnonic devices becomes one major concern as it determines how the signals are being processed. Thus, in present work, particular attention is paid to the interaction of vortex with unidirectional propagation of spin wave (SW), and we address three crucial problems: 1). How is the propagating SW being scattered? 2). How does vortex react to the propagating SW? 3). How can we take advantage of such SW-vortex interaction? Particularly, for vortex with stabilized winding number +1/2 or -1/2 (i.e., magnetization at disk edge keeps tangent, hence avoidance of surface magnetic charges), it could be viewed as half a Bloch-type skyrmion as it covers half of the sphere in order parameter space $V = \mathbb{S}^2$. The topologically nontrivial texture is thus promising to yield magnon Hall effect as in skyrmions [48], which is of great interest and importance.

To reveal the interaction details, the underlying magnetization dynamics that's governed by Landau-Lifshitz-Gilbert equation [49,50]

$$\frac{\partial \boldsymbol{m}}{\partial t} = -\gamma \boldsymbol{m} \times \boldsymbol{h}_{eff} + \alpha \boldsymbol{m} \times \frac{\partial \boldsymbol{m}}{\partial t}, \quad (1)$$

(here $\gamma$ is the gyromagnetic ratio; $\boldsymbol{h}_{eff}$ is the effective field that includes exchange interaction, dipole-dipole interaction, magnetocrystalline anisotropy and Zeeman energy; $\alpha$ is the Gilbert damping constant), is majorly numerically solved based on Mumax$^3$ code [51], whereas the related material parameters are adopted from permalloy (i.e. saturation magnetization $M_S = 8.6 \times 10^5$ A m$^{-1}$, exchange stiffness $A = 1.3 \times 10^{-11}$ J m$^{-1}$) and the unit cell is chosen to be 4×4×5 nm$^3$ [52-54]. During the simulation, the Gilbert damping constant is set to be $\alpha = 0.001$ to ensure a long-distance propagation of the SW beams [55].

This paper is organized as follows. In Sec. II, by tracing the trajectories of Gaussian beams of SW through the magnetic texture, we reveal strong unilateral skew scattering and relatively weak symmetric side deflection of SW. Based on the collective description of the interaction in Lagrangian frame, the skew scattering is attributed to the topology-induced emergent field inside the core region. The symmetric side deflection, on the other hand, can be explained in terms of the local energy density distribution of the vortex. In Sec. III, from the fast Fourier transformations of the temporal variations of vortex-core position, we identify two modes of gyrations: a circular mode with fixed sub-GHz frequency and an ellipse-type mode with frequency same as that of the propagating SW, which can be ascribed to permanent and periodic magnon spin-transfer torque effects, respectively. Sec. IV focuses on our proposal of vortex-based SW valve, in which we achieve tunable phase shift of SW beside successful valving. Additionally, the phase shift is determined to be a monotone decreasing function of degree and dimensional span of the inhomogeneity of the contouring texture. Finally, a brief summary is given in Sec. V.

## II. Scatterings of spin wave by vortex

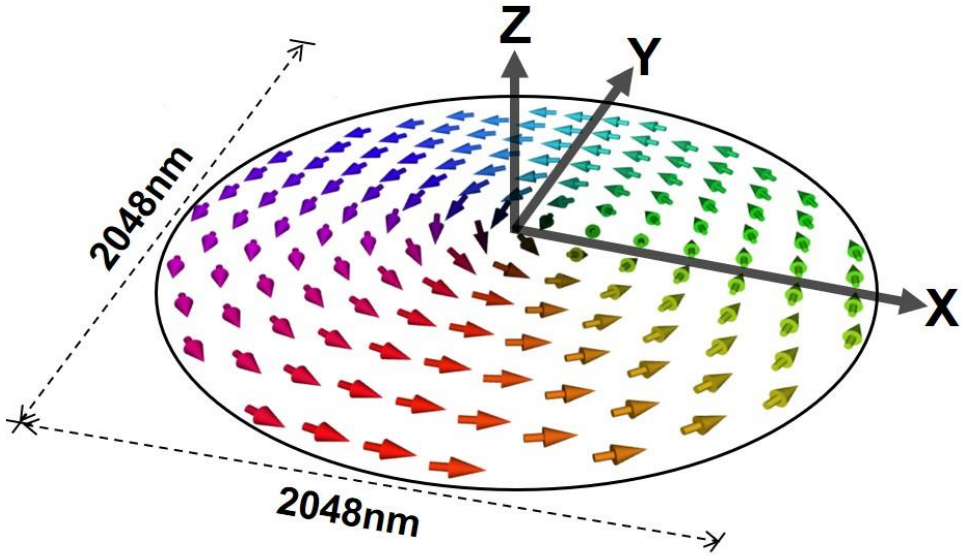

FIG. 1. Schematics of magnetic vortex in disk

To start with, for the idealization and simplicity of model, we construct a single-layer disk with a diameter of 2048 nm, in which a standard vortex is initially stabilized with its core located at the center, see Fig. 1. The vortex core is spontaneously regularized to have an out-of-plane polarity instead of a singular point defect, hence the so-called Heisenberg vortex. To get a visualization on how SWs are being scattered by the vortex, a straightforward method is to trace their travelling trajectories, which can be done using Gaussian beams [55,56]. Here, we prepare multiple SW Gaussian beams at different vertical locations on the left side so that a thorough image of scattering can be obtained. Each beam is excited by applying a sinusoidal monochromatic microwave field $\boldsymbol{H}_{ac} = h_0 \sin(\omega t) \hat{z}$, where $\omega/2\pi = 100$ GHz in a narrow rectangular region with its long side parallel to the wave front that is expected to move forward to the right (i.e., +x direction). The field amplitude $h_0$ has a Gaussian distribution along y axis with its maximum set to be $h_{max} = 5$ kOe. In order to minimize the reflection of SW from the disk edge, an absorbing boundary condition with increased damping constant is adopted for outer-ring region, see shadowed area in Fig. 2.

The tracked trajectories of the Gaussian beams have been highlighted using solid or dashed curves in Fig. 2. For SW that passes through the central vortex core (VC) region, one clearly finds strong unilateral and sectorial scattering of the SW (see the black dashed curves that outline a spectrum of scatterings of a single Gaussian beam right across the center of the vortex), giving rise to topological magnon Hall effect [57,58,59,60]. Notably, the direction of such scattering definitely depends on the polarity of the VC, in which -1 (+1) gives rise to downward (upward) scattering. On the other hand, for SWs that bypass the VC region, a relatively weak and symmetric side deflection is observed (see the white solid curves), and the beam trajectory depends on the impact parameter (i.e., the vertical shift from VC). In general, those beams with smaller impact parameters are more strongly deflected. Surprisingly, a small and temporal fluctuation may occur when SWs enter the close surroundings of VC (see the central segments of the two white solid curves nearest to the core). For large enough impact parameters, the deflection can be neglected. Through comparison between Fig. 2(a) and 2(b) of 180-degree x-axis rotation, one may further learn that the chirality affects neither the deflection outside the VC region nor the strong scattering inside the VC region. It is worth mentioning that according to our observation, SWs with reduced frequency and amplitude lead to similar scattering behaviors.

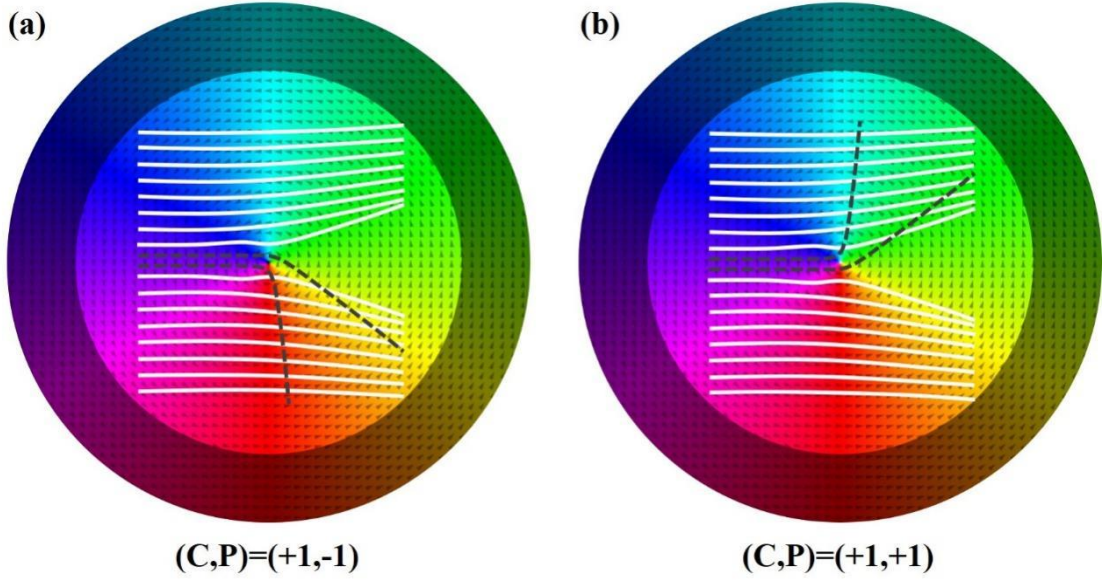

FIG. 2. Schematics of propagating SWs across (a) magnetic vortex with chirality +1 and polarity -1 (i.e., $(\mathbf{C},\mathbf{P}) = (+\mathbf{1},-\mathbf{1})$) and (b) magnetic vortex with chirality +1 and polarity +1 (i.e., $(\mathbf{C},\mathbf{P}) = (+\mathbf{1},+\mathbf{1})$) within disks (diameter & thickness: 2048 nm & 5nm). The outer shadowed ring (width: 256 nm) are implemented with an absorbing boundary condition that gradually increases damping constant $\alpha$ as approaching the edge. SW Gaussian beams are excited on the left about 512 nm away from core, and are vertically evenly spaced with interval 64 nm. The solid white lines represent the propagating trajectories of the Gaussian beams through the vortex. The dashed black lines outline the sectorial scattering of a Gaussian beam across the center of the vortex.

In order to explain the strong sectorial scattering inside VC, here we follow the general dynamics of spin wave and magnetic texture formulated in a Lagrangian frame [48]. For the Heisenberg vortex that can be embedded in a Bloch-type skyrmion, by invoking the Euler-Lagrangian rule, employing minimal collective coordinates $X_\mu = (X, Y)$ and neglecting texture corrections due to existence of local SW excitations, one can arrive at equations of motion in real space as:

$$\dot{X} \times B^0 \hat{z} = -\rho \dot{x} \times \vec{\mathcal{B}}, \qquad (2a)$$

$$m_{sw}\ddot{x} + \dot{x} \times b^0 \hat{z} = \dot{X} \times \vec{\mathcal{B}}, \qquad (2b)$$

where $m_{sw}$ is the normalized effective mass of the excitation of SW packet, $\rho$ is the total SW intensity, $X$ and $x$ denote the central positions of the texture and the wave packet,

$$b^0 = \boldsymbol{m}_0 \cdot \left(\frac{\partial \boldsymbol{m}_0}{\partial x} \times \frac{\partial \boldsymbol{m}_0}{\partial y}\right), \qquad (3)$$

is the effective magnetic field that corresponds to the magnetic topology of the texture,

$$B^0 = B^0_{xy} = \int \boldsymbol{m}_0 \cdot \left(\frac{\partial \boldsymbol{m}_0}{\partial x} \times \frac{\partial \boldsymbol{m}_0}{\partial y}\right) dx dy = 4\pi Q, \qquad (4)$$

is the topological charge-associated magnetic field (here $Q$ represents topological charge) that can in turn deflect the texture itself, $\vec{\mathfrak{B}} = b^0\hat{z}$ formulates the total mutual magnetic field spanning both parameter spaces $\{X_\mu, x_i\}$ in which the texture and SW interact with each other (note that as the DMI is not involved in the vortex, contributions from the effective magnetic field $b^D$ have been omitted). It is worth noting that the right sides of Eq. (2a) and Eq. (2b) indicate the magnonic spin transfer torque exerted on the texture and the texture-induced motive force acting on the SW [61,62], respectively. According to Eq. (2b), obviously, a transverse velocity can be expected to arise for a SW travelling upon an inhomogeneous texture, which defines the skew scattering for SWs [48].

For a typical Bloch-type skyrmion profile, the spatial distribution of field $\vec{\mathfrak{B}}$ is rotationally symmetric, and thus its value is only a function of radius. Notably, there exists a narrow annulus within (beyond) which the field is of positive (negative) value. Subsequently, both upward and downward scatterings of SW occur [48]. In current study, since the Heisenberg vortex is equivalently half of the skyrmion, the effective magnetic field is thus expected to remain either all positive or all negative. In Fig. 3, we show the 2D topological charge density ( $q = (1/4\pi)\boldsymbol{m} \cdot (\partial_x \boldsymbol{m} \times \partial_y \boldsymbol{m})$ ) distribution, which is in proportion to the effective magnetic field (or emergent field [63]) $b^0$. Non-zero $q$ does arise but within only the narrow VC region described in terms of the polarization or $m_z$ profile. The $q$ (hence emergent field $b^0$) changes rapidly as approaching the core center. However, the sign of $q$ indeed remain unchanged and depends definitively on the polarity, as expected. Due to the effective Lorentz force $F_j = v_i b_{ij}^0$ which SW packets would experience as they pass through the VC, the unilateral skew scattering corresponding to the Hall effect of magnons inside core region naturally emerges.

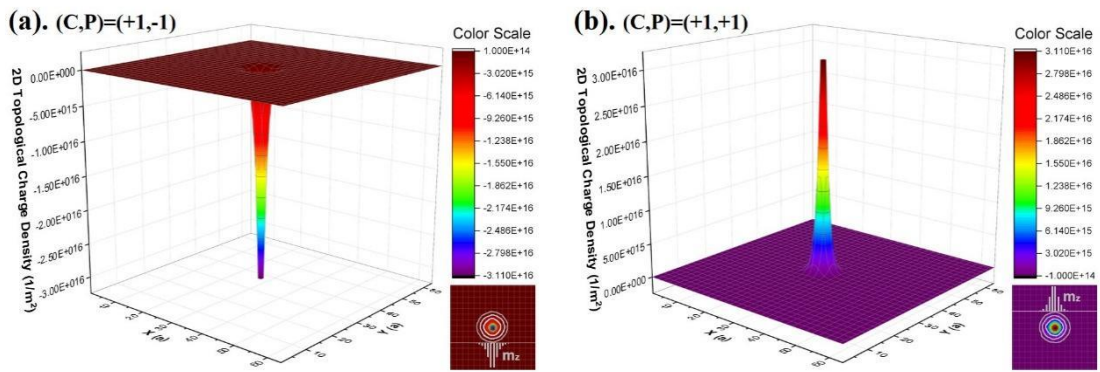

FIG. 3. Distribution of 2D topological charge density ($q$) in (a) magnetic vortex with $(\mathbf{C}, \mathbf{P}) = (+1, -1)$ and (b) magnetic vortex with $(\mathbf{C}, \mathbf{P}) = (+1, +1)$. Insets at the lower right corners emphasize the small area of dominant $q$ within radius of ~8a (i.e., a=4 nm), which is comparative to the non-zero $m_z$ profile of core.

To be worth raising, as shown in Fig. 4, we also observe that under the propagation of SW beam right across the central VC region, the core itself is pulled to the left while undergoing upward (downward) movement for $P = -1$ ($P = +1$) due to the reaction force out of the skew scattering (recall that the VC can be deflected by its own topological field $B^0$), which consists with calculations based on Eq. (2a). The oscillatory manner is believed to attribute to the process of periodic magnonic spin-transfer torque, as will be explained in Sec. III.

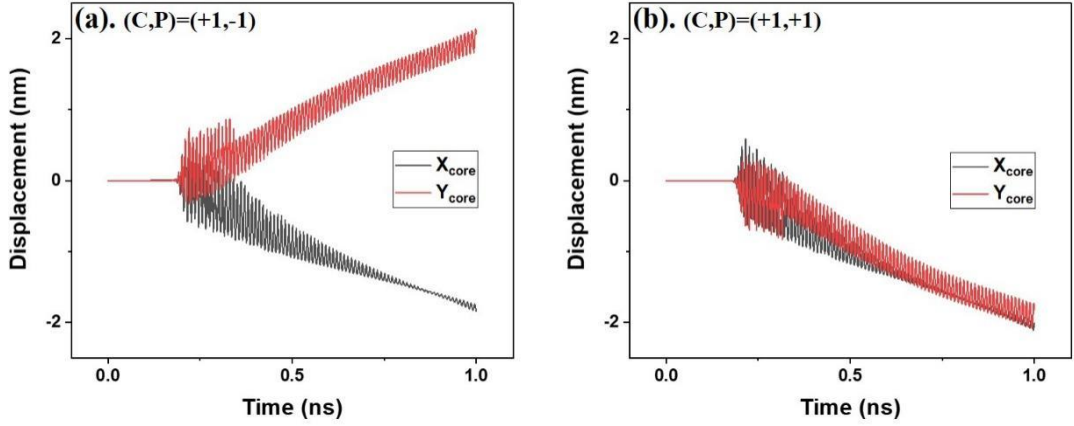

FIG. 4. 2-dimension movements of VC under propagation of SW beam through central VC region in (a) magnetic vortex with $(\mathbf{C}, \mathbf{P}) = (+\mathbf{1}, -\mathbf{1})$ and (b) magnetic vortex with $(\mathbf{C}, \mathbf{P}) = (+\mathbf{1}, +\mathbf{1})$.

In order to understand the symmetric side deflection of SW outside VC, one may further examine the linearization of the Landau-Lifshitz-Gilbert equation under perturbation. In many cases including the above-mentioned Bloch-type skyrmion, the linearized LLG equation can be recast approximately into an effective Schrödinger equation as [64-66]:

$$i\hbar(1+i\alpha)\frac{\partial \Psi}{\partial t} = \left[\frac{1}{2\mu}(-i\hbar\nabla - \mathbb{A})^2 + \mathcal{V}\right]\Psi, \tag{5}$$

where $\Psi = m_\theta - im_\phi$ ($m_\theta$, $m_\phi$ represent the SW excitations at polar and azimuthal angles of $\hat{\boldsymbol{m}}_0$, $\hat{\boldsymbol{m}}_0$ denotes static magnetization), $\mu = \hbar/2\gamma A$ is the effective SW mass, $\mathbb{A}(\boldsymbol{r})$ and $\mathcal{V}(\boldsymbol{r})$ are respectively the spatially varying effective vector and scalar potentials determined by the static magnetization profile. Considering the existence of the Gilbert damping constant on the left hence no probability conservation, propagation of magnons in $\mathbb{A}(\boldsymbol{r})$ and $\mathcal{V}(\boldsymbol{r})$ is equivalent to electrons moving in certain vector potential and scalar potential. The vector potential (note that the curl of it $\nabla \times \mathbb{A}(\boldsymbol{r})$ [57,66] gives rise to effective magnetic field) and the corresponding effective magnetic field of vortex hence the skew scattering/topological Hall effect of SW [57,67] have been reasonably discussed already. The scalar potential $\mathcal{V}(\boldsymbol{r})$ stemming from conventional texture inhomogeneity [64] however has been disregarded so far.

In Fig. 5(a), we show the radical distribution of magnetic energy that's expected to proportion the scalar potential [66] within diameter of 64a in VC with respect to the $m_x$ and $m_y$ profiles for case $(\mathbf{C}, \mathbf{P}) = (+\mathbf{1}, -\mathbf{1})$ (note that for simplicity, we will focus on only magnetic vortex with $(\mathbf{C}, \mathbf{P}) = (+\mathbf{1}, -\mathbf{1})$ in the rest of the paper). The total energy is mainly contributed by the exchange energy (see Fig. 5(b)) and finite energy density arises within only limited area near core with a dominant peak at the very center. Nevertheless, through comparison between the total energy density ($E_{dens\_total}$) distribution and the $q$ distribution in $ln$ scale (see Fig. 5(c)), the former still turns out to be a lot more extensive in radical dimension. As seen, the $q$ is highly confined within merely 16a in diameter. Such difference between the $E_{dens\_total}$ and $q$ in their active area could be responsible for the regionalization of the two scattering behaviors shown in Fig. 2.

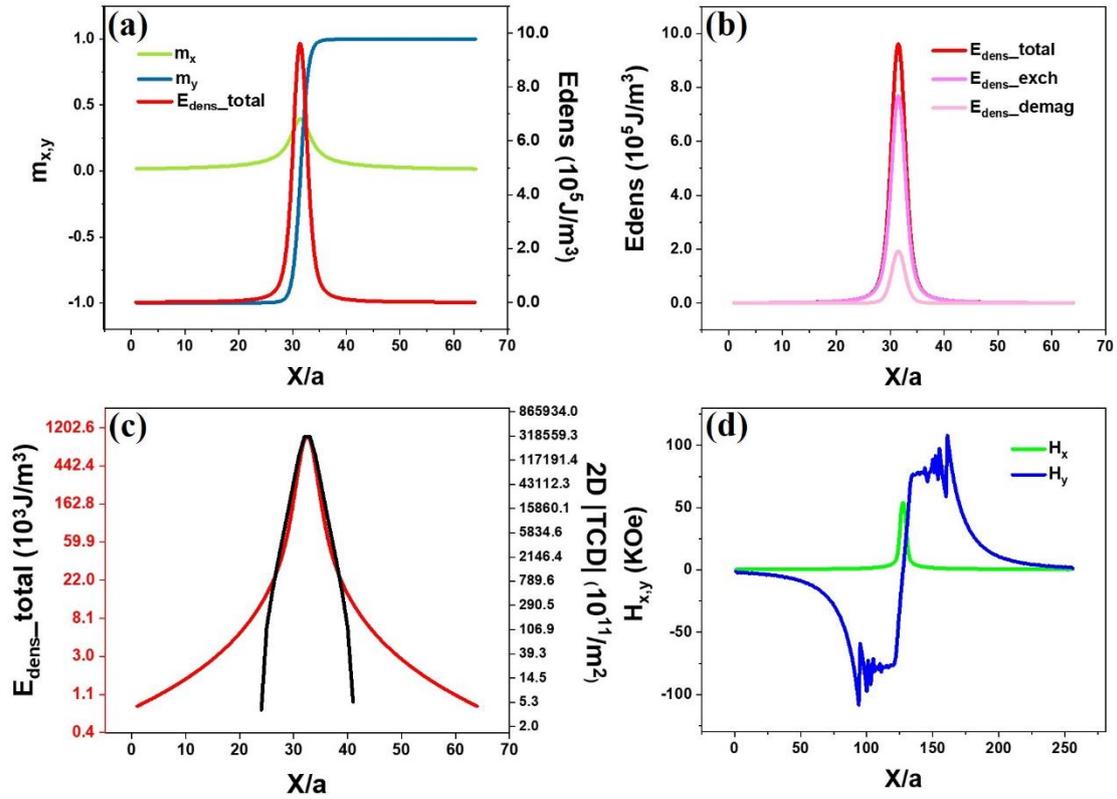

FIG. 5. Comparison between total energy density distribution and (a) magnetization distribution (here $m_i = M_i/M_S$) (b) the components of exchange energy density and demagnetization energy density distributions (c) the absolute $q$ distribution (in $ln$ scale) over circle diameter of 64a (i.e., 256 nm) across the VC along x axis, based on $(\mathbf{C}, \mathbf{P}) = (+\mathbf{1}, -\mathbf{1})$. (d) The corresponding internal-field distributions of $H_x$ and $H_y$ over circle diameter of 256a (i.e., 1024 nm) across the VC. Considering the rotational symmetry of the vortex, relevant characters are plotted with respect to x axis only.

It is well known that the partial derivative of total energy density with respect to local magnetization corresponds to the effective field $\mathbf{H}_{eff}$ (recall that $\mathbf{H}_{eff} \propto \partial E_{dens}/\partial \mathbf{m}$). The internal field that's due to exchange and dipolar interactions thus has been subsequently determined over circle diameter of 256a across the VC, as plotted in Fig. 5(d). As seen, component $H_x$ ($H_y$) possesses the same symmetry as component $m_x$ ($m_y$). However, $H_y$ clearly exhibits enhanced inhomogeneity compared with $m_y$. By sweeping the $H_x(x)$ and $H_y(x)$ distributions around z axis (recall the rotational symmetry of the vortex) while disregarding their signs, 3D absolute field profiles can be obtained as shown in Fig. 6, from which formation of potential barriers and wells can be observed. In viewing of the fact that the radical field [see Fig. 6(a)] is highly confined within only the narrow area near core and is insignificant compared with the azimuthal field [see Fig. 6(b)], we expect a potential basin near the center of the disk and a highland within interval R(8a, 34a), but both are with certain fluctuations. Beyond radius of 34a, the potential decreases monotonously as approaching the disk edge. Such crater-shaped potential rationalizes the symmetric side deflection outside VC region, whereas the anomalous fluctuations observed in SW trajectories could be explained in terms of correspondingly potential fluctuations.

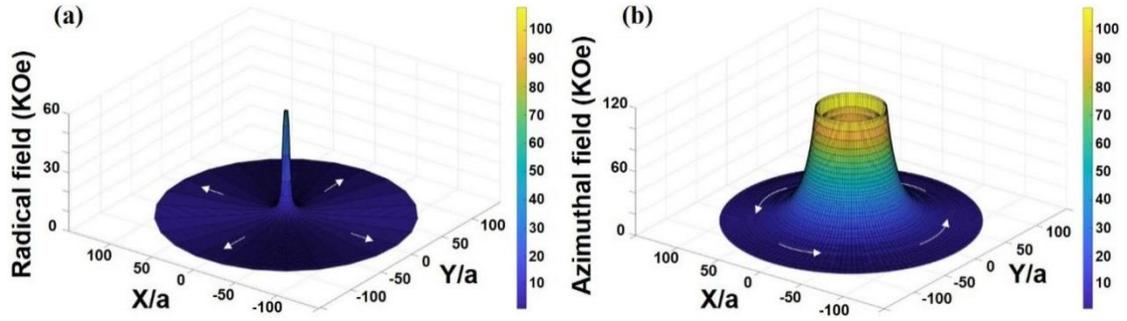

FIG. 6. (a) 3D absolute-value profile of the internal radical field $H_{radical}(r) = |H_x(x)|_{x=r}|$. (b) 3D absolute-value profile of the internal azimuthal field $H_{azimuthal}(r) = |H_y(x)|_{x=r}|$.

As $\mathbf{H}_{eff} \propto \partial E_{dens}/\partial \mathbf{m}$, it is natural to predict that changed $E_{dens\_total}$ and/or magnetization field $\mathbf{m}_i$ shall lead to different scatterings of spin wave. In Fig. 7(a) and 7(b), we obtain successful modulation on the scattering performance through simulations of spin systems with enlarged size or modified geometry. By repeating the Gaussian beam simulations, the beam trajectories have been traced using solid black lines, in contrast to the original trajectories (see the dashed red lines) reproduced from Fig. 2(a). Misalignments of the trajectories are clearly seen. More precisely, in present two cases, spin waves all become less deflected. Hypothetically, in order to get stronger deflections, we reckon that one may try out spin systems with reduced size or increased exchange stiffness.

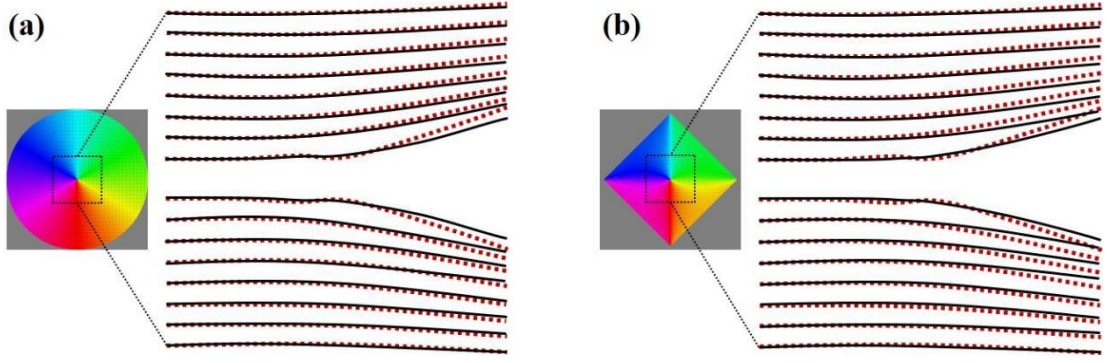

FIG. 7. Out-of-core trajectories (see the black solid lines) of Gaussian beams in (a) magnetic vortex within disk of diameter 3072 nm. (b) magnetic vortex within square plate with edge length 2048 nm. The dashed red lines mark the trajectories of out-of-core side deflections in the original vortex within disk of diameter 2048 nm.

### III.    Driven translational modes of vortex core

In this section, we focus on the antithesis of Sec. II—the driven motion of VC by propagating SWs, which represents the other type of spin excitations (namely translational or gyrotropic modes) existing in magnetic vortex system [30,40,68]. The upper left (lower left) movement displayed in Fig. 4(a) (Fig. 4(b)) only reflect the transient behavior of core under a narrow Gaussian beam. Persistent mode of motion when subjected to constant pumping of SWs however remain unknown. To crack the problem, herein we generate plane waves on the left with sufficiently larger widths compared with the size of VC. In Fig. 8, we show several representative motion patterns of core under constant pumping of SWs in different amplitudes and frequencies. Note that the core trajectories are traced in terms of the guiding center (X, Y) of core, where $X = \int xq_t dxdy / \int q_t dxdy$, $Y = \int yq_t dxdy / \int q_t dxdy$ [69]. The patterns exhibit great complexity and largely differ from each other. Nonetheless, we find accordant sense of rotation among them given the same **C & P** condition. The sense of rotation follows a simple rule as tabulated in Tab. 1, which consists with Thiele's theory [70] and early experimental findings [68]. In present case where $(\mathbf{C},\mathbf{P}) = (+\mathbf{1},-\mathbf{1})$, the motions generally rotate clockwise (CW).

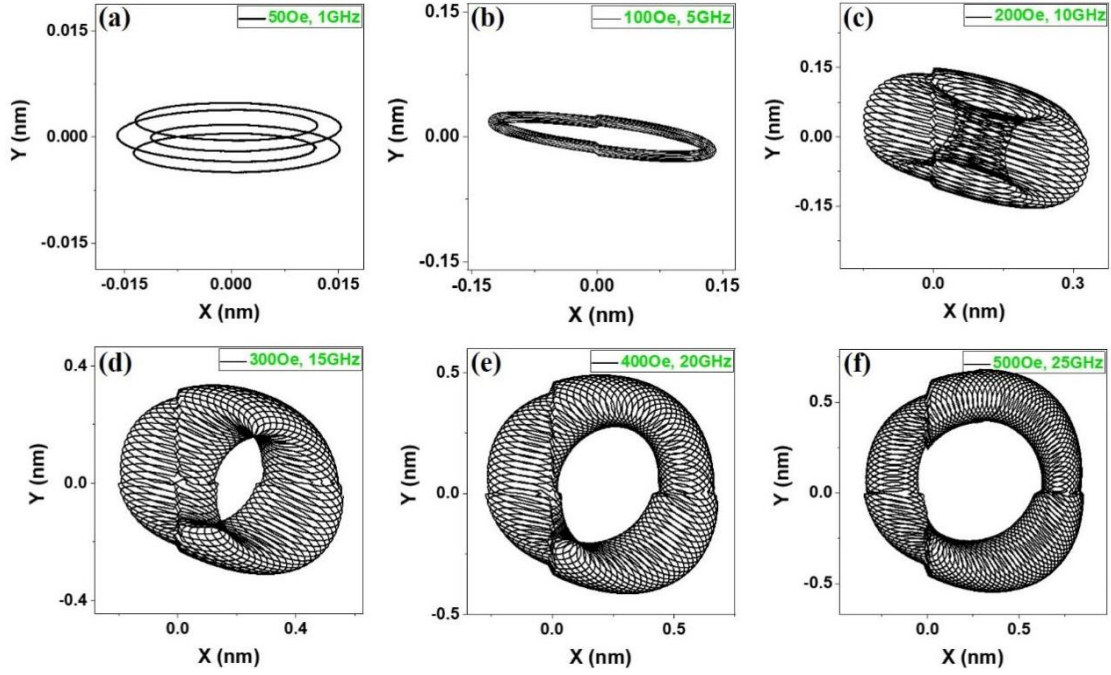

FIG. 8. Driven-motion trajectories of VC under plane waves excited by $\boldsymbol{H}_{ac} = h_0 \sin(\omega t)\hat{z}$) on the left about 512 nm away from core center with (a) $h_0 = 50$ Oe, $\omega/2\pi = 1$ GHz (b) $h_0 = 100$ Oe, $\omega/2\pi = 5$ GHz (c) $h_0 = 200$ Oe, $\omega/2\pi = 10$ GHz (d) $h_0 = 300$ Oe, $\omega/2\pi = 15$ GHz (e) $h_0 = 400$ Oe, $\omega/2\pi = 20$ GHz (f) $h_0 = 500$ Oe, $\omega/2\pi = 25$ GHz.

TAB. 1. Rotation senses of core motion with respect to **C** & **P**.

| C \ P | -1 | +1 |
|---|---|---|
| -1 | CW | CCW |
| +1 | CW | CCW |

As known, deviation of VC from equilibrium induces magnetic charges and largely affects the magnetostatic energy of the vortex system [71,72]. Due to the bowl-shaped potential well bound to the vortex, Eq. (2) is no longer sufficient to describe motion of core in disk. To comprehend various patterns in Fig. 8, we carry out Fast Fourier Transform (FFT) analysis to the temporal variations of core position over considerable periods. The obtained FFT power spectra in frequency domain reveal two prominent peaks: an ever-present low frequency of 0.24 GHz and a high frequency exactly same as the pumping SW, see Fig. 9. The FFT intensities in different cases largely differ. However, in each case, for the 0.24 GHz, X, Y components are of the same amplitude, which suggests a circular motion of VC. For the high frequency though, FFTs of the two components appear to be different, hence an elliptical motion of the core, which is similar to observations in the study of vortex in cross-structure waveguides [29]. A close look at the FFT spectra further reveals a series of minor peaks, see for example Fig. 10(a) that corresponds to the amplified image of Fig. 9(f) near 25 GHz. These minor peaks may arise from the couplings of the two modes of motion in view of the fact that their frequencies can be expressed as $25 \pm 0.24n$ GHz where $n$ denotes an integer. The intensities of such mixed modes depend on the amplitudes of the original circular and elliptical modes. As seen in Fig. 10(b) that corresponds to the amplified image of Fig. 9(b) near 5 GHz (note that the amplitude of peak 0.24 GHz becomes insignificant in Fig. 9(b)), the first-order $25 \pm 0.24$ GHz appear to be extremely small and the higher orders are barely observed.

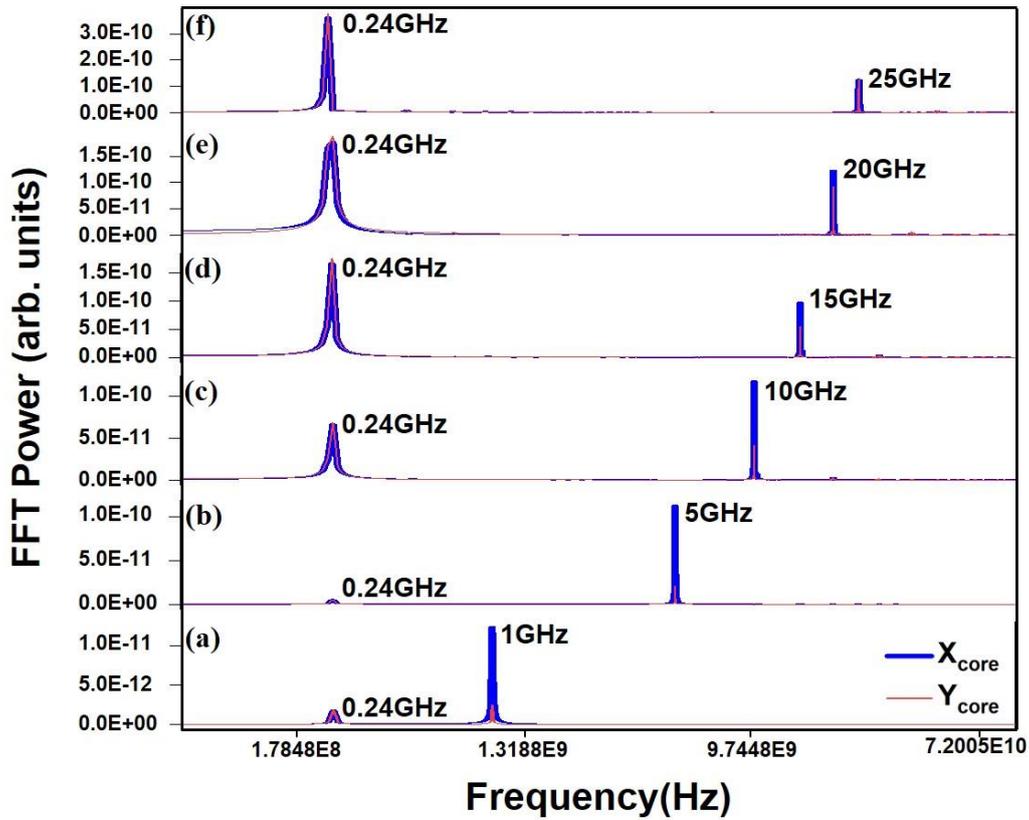

FIG. 9. FFT spectra of the driven motions of VC corresponding to Fig. 8 in alphabetical order. The FFT of the X and Y components are respectively labelled with blue and red curves. The scale of x axis is of $ln$ type.

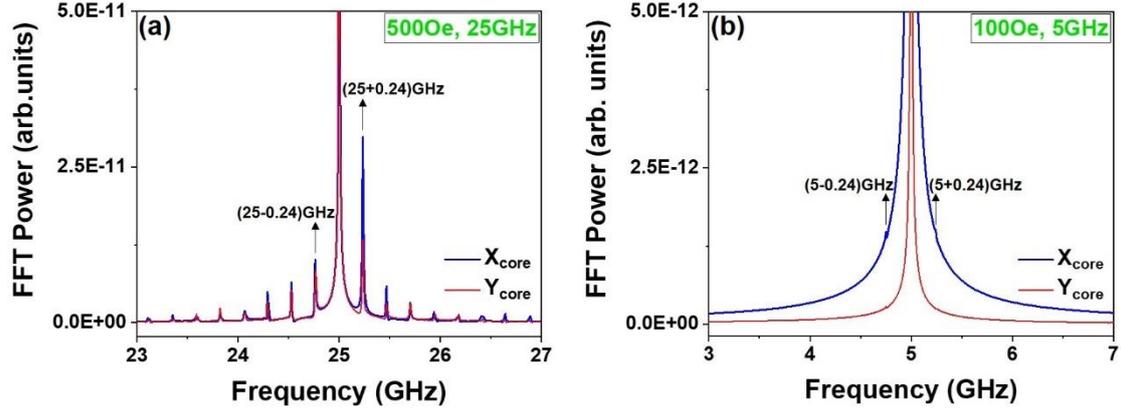

FIG. 10. (a) Zoomed-in FFT spectrum of the driven motion of VC near 25 GHz for ($h_0 = 500$ Oe, $\omega/2\pi = 25$ GHz). (b) Zoomed-in FFT spectrum of the driven motion of VC near 5 GHz for ($h_0 = 100$ Oe, $\omega/2\pi = 5$ GHz). The FFT of the X and Y components are respectively labelled with blue and red curves.

A visualization of the two basic modes of motion is given in Fig. 11, where one can find varied amplitude of the circular mode as well as varied amplitude, ellipticity (i.e., long-to-short axial ratio) and tilting angle of the elliptical mode, with respect to $h_0$ and $\omega$. In addition to the vivid orbiting motions of the two modes, deviations of motion centers from disk center are also appended, as labeled by DEV. Notably, in all cases, the motion centers shift to the upper right. The upward deviation, needless to say, can be attributed to the strong skew scattering of SW. The forward deviation however lacks of explanation. Considering the common phenomena of SW reflection by the many magnetic solitons (e.g., domain wall) due to rise of dynamic stray field [31,73-75], we compare the arrived SWs at a spot in front of the disk center in cases in absence of VC and in presence of VC. By subtracting the SW of former case from that of the latter, we observe clear back reflection of SW by the VC (see for instance the red curve in Fig. 12 that is calculated based on case under $h_0 = 500$ Oe and $\omega/2\pi = 25$ GHz), which suggests a process of linear momentum transfer torque (LMTT) [73,74]. Thus far, one may conclude three scattering behaviors of the propagating SW in vortex: Skew scattering, side deflection and back reflection. On the other hand, the skew scattering and back reflection of SW together yield steady reaction forces on the core and can further balance with the restoring force provided by the bowl-shaped potential well bound to the vortex.

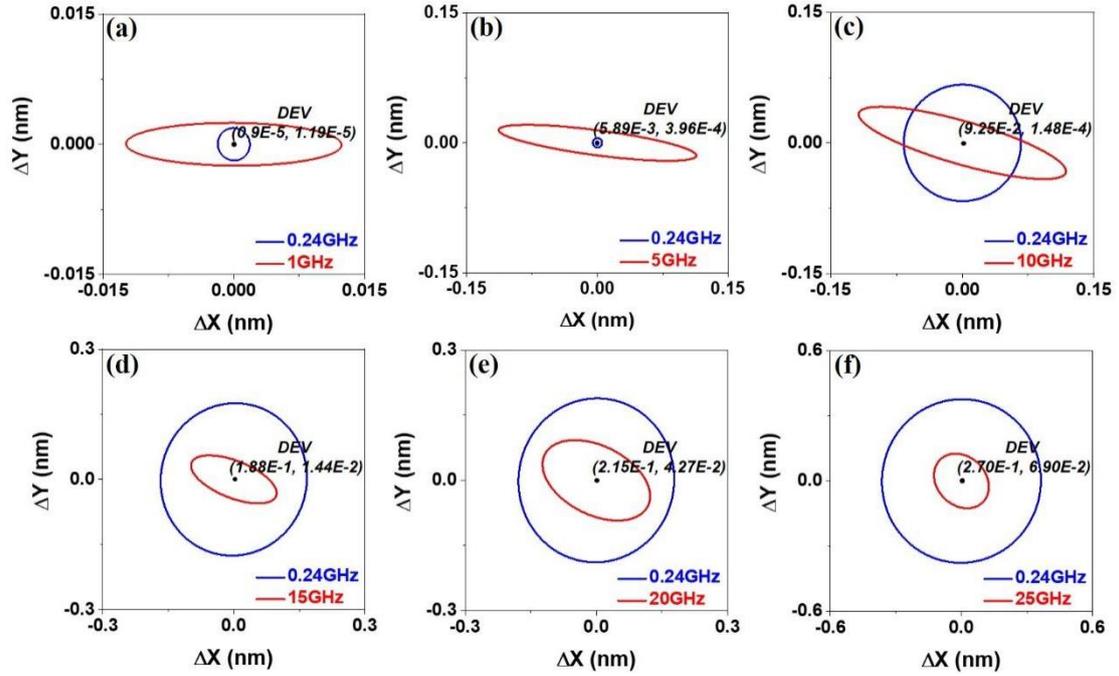

FIG. 11. Corresponding trajectories of the individual FFT peaks in Fig. 9. Blue and red curves respectively sketch the component circular motion and elliptical motion of VC. DEV denotes the in-plane deviation of motion center from disk center.

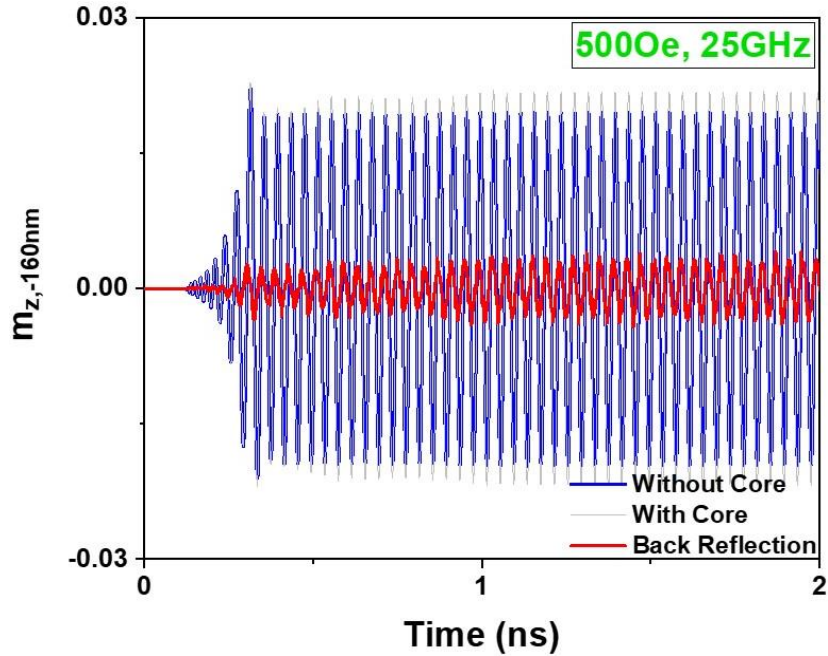

FIG. 12. Normalized magnetization component $m_z$ at 160 nm left away from vortex core as a function of time under SW excitation with $h_0 = 500$ Oe and $\omega/2\pi = 25$ GHz. The blue curve denotes the case in absence of VC, which is simulated based on uniform magnetization since the spin textures in standard vortex along radical direction within certain lateral range can be approximately viewed as uniform. The grey curve denotes the case in presence of VC. The highly overlapped magnetization

oscillations in the early stage of 0.5 ns between blue curve and grey curve evidence the effectiveness of uniform magnetization as an alternative in simulating SW propagation in absence of VC. The red curve denotes the back reflection of spin wave by the VC.

To understand the motion modes of VC, further simulation under SW of 0.24 GHz (as an illustration, here condition of $h_0 = 50\text{ Oe}$, $\omega/2\pi = 0.24\text{ GHz}$ is used) is priorly performed with recorded trajectory shown in Fig. 13(a). As one may expect, excitation of the circular mode at 0.24 GHz is continuously observed. However, the elliptical mode disappears, see the FFT spectrum in Fig. 13 (b). Fig. 13 (c) and 13(d) display the case where the core undergoes naturally damped spiral gyration [76] with the intrinsic frequency that happens to be 0.24 GHz. The coincident peaks in Fig. 13(b) and 13(d) suggests that the circular mode in fact corresponds to undamped 2D harmonic oscillations of the VC.

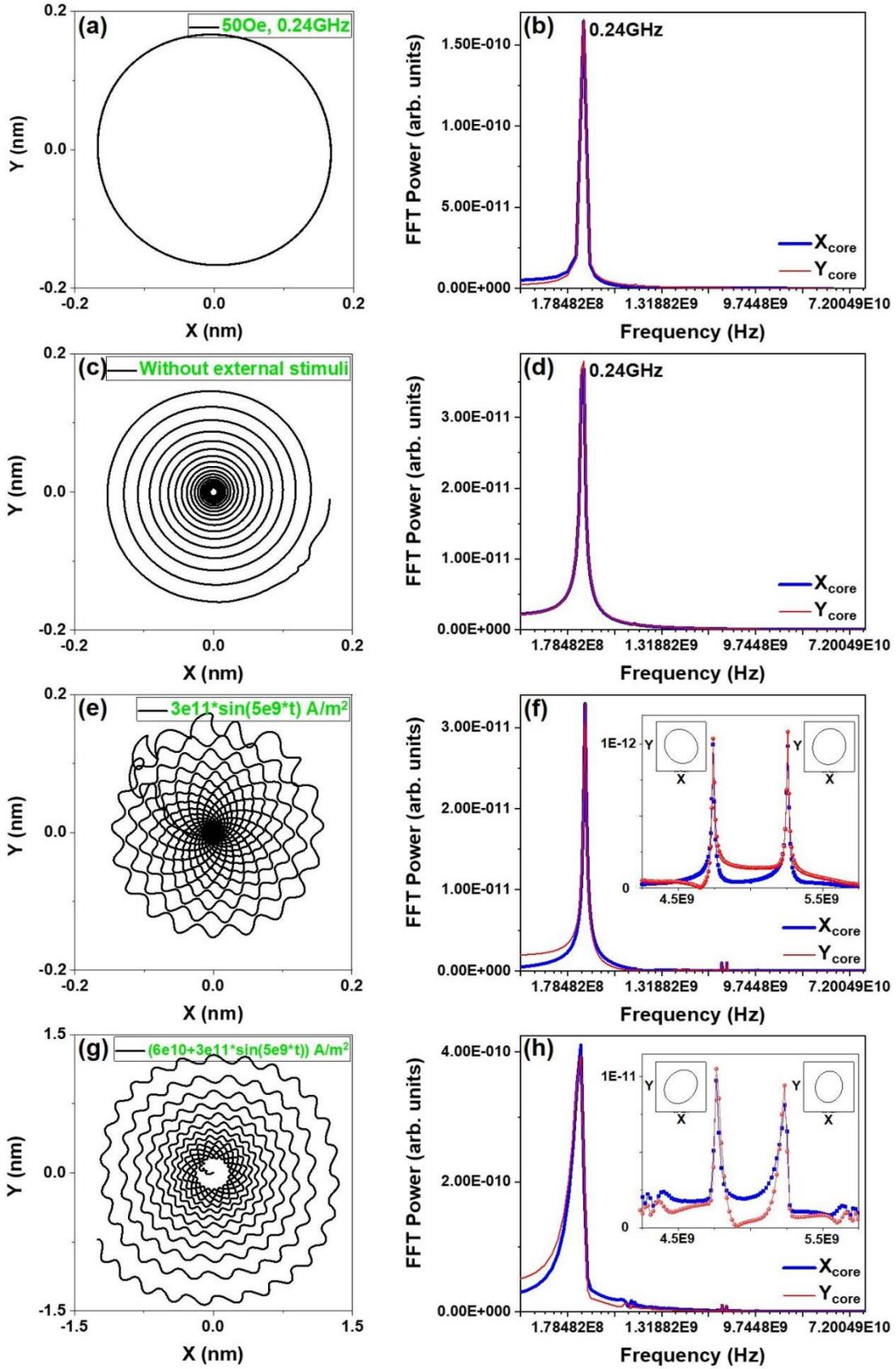

FIG. 13. Motion trajectory of VC by (a) SW based on $h_0 = 50$ Oe, $\omega/2\pi = 0.24$ GHz (c) no external stimuli (e) spin polarized current of $3 \times 10^{11} \times \sin(5 \times 10^9 \times t)\ A \cdot m^{-2}$ (g) spin polarized current of $(6 \times 10^{10} + 3 \times 10^{11} \times \sin(5 \times 10^9 \times t))\ A \cdot m^{-2}$.

Note that in (c), (e) and (g), the cores are initially displaced by a field pulse. The current polarization $P$, the Slonczewski parameter $\Lambda$ and Slonczewski secondairy STT term $\varepsilon'$ in Slonczewski STT model are respectively set to be 0.2, 1 and 0, whereas the unit magnetization vector in fixed layer is (0, 0, 1). (b), (d), (f) and (h) are respectively FFT spectra of (a), (c), (e) and (g) showing averaged FFT power. Insets in (f) and (h) demonstrate two mixed modes near 5 GHz, in which ellipticity can be found.

To explain how such undamped harmonic oscillations can be sustained, we follow Thiele's formulism [70] that assumes a rigid vortex profile with its internal magnetization characterized solely by VC position $\boldsymbol{X}$ (i.e., $\boldsymbol{m}(\boldsymbol{r}) = \boldsymbol{m_0}(\boldsymbol{r} - \boldsymbol{X})$, where $\boldsymbol{r}$ denotes space coordinate and $\boldsymbol{m_0}$ is the initial magnetization distribution when the core locates at origin). Taken into account of the spin angular momentum transfer effect, the Thiele equation writes:

$$\mathcal{G}_{ij}\dot{X}_j = -\partial_i \mathcal{W} + \mathcal{D}_{ij}\dot{X}_j + \mathcal{F}^i_{ST}, \tag{6}$$

where the left-side term represents the gyrotropic force, the first, second and third terms on the right side are respectively restoring force (here $\mathcal{W}$ denotes total magnetic energy), dissipation force and spin-torque force. Components of the gyrotensor $\hat{\mathcal{G}}$ and dissipation tensor $\hat{\mathcal{D}}$ can be expressed as follows [77]:

$$\mathcal{G}_{ij}(\boldsymbol{X}) = \frac{M_S}{\gamma} \int d^3\boldsymbol{r}(\partial_i \boldsymbol{m} \times \partial_j \boldsymbol{m}) \cdot \boldsymbol{m}, \tag{7}$$

$$\mathcal{D}_{ij}(\boldsymbol{X}) = -\alpha_{LLG} \frac{M_S}{\gamma} \int d^3\boldsymbol{r}(\partial_i \boldsymbol{m} \times \partial_j \boldsymbol{m}). \tag{8}$$

Notably, the gyrotropic force and restoring force are of radical direction, whereas the dissipation force is perpendicular to VC displacement [77,78]. Based on the Thiele equation, one immediately learns that in order to sustain an undamped circular motion in which $\dot{\boldsymbol{X}} = \boldsymbol{\omega} \times \boldsymbol{X}$ holds, the dissipation force must be compensated by a steady spin-torque force. Spin polarized current (SPC) has been shown to be able to counteract dissipation and successfully maintain circular motion of VC [77,78]. Specifically, when a SPC flows perpendicularly to the disk plane, the concomitant spin-transfer (ST) torque term $\boldsymbol{\tau}_{SPC} = \sigma J \boldsymbol{m} \times (\boldsymbol{m} \times \boldsymbol{P})$ gives rise to a spin-torque force [79,80] $\mathcal{F}^i_{ST}(\boldsymbol{X}) = M_S \sigma J P \int d^3\boldsymbol{r}(\boldsymbol{m} \times \partial_i \boldsymbol{m})$ where $\sigma = \hbar\eta/(2L|e|M_S)$, $\eta$ is current spin polarization, $L$ is free layer thickness, $J$ is current density, $\boldsymbol{P} = P\hat{\boldsymbol{z}}$ represents unit vector of polarizer magnetization ($P = \pm 1$). Obviously, SPC excites VC motion when the electrons bring a magnetic moment from the polarizer to free layer opposite to the core polarization (i.e., the overall ST torque $\boldsymbol{\tau}_{SPC}$ tends to reduce the degree of polarization of the core). Like SPC, a propagating SW through core can be expected to yield similar effect. Note that propagation of SW inevitably causes precession of core magnetization, which naturally involves a process of spin-transfer torque (STT). In Fig. 14(a) and 14(b), we record the core magnetization components subjected to SW of 0.24 GHz. A gradual enlarging of the precession of core is clearly companied with a monotonous reduction in net core magnetization along z axis (see the red double-headed arrow), evidencing considerable magnonic STT $\boldsymbol{\tau}_{SW}$. Based on such magnonic STT term, sustained circular motion of VC becomes a natural consequence when under constant SW propagation. As the propagating SW gets stronger, one may further expect an increasing orbiting radius (we will show this in later content). Interestingly, the ability of SW in driving such intrinsic circular motion also extends to general frequencies, as already demonstrated in Fig. 9 which includes the circular motion as a component mode in FFT spectrum. According to our theory, one thus expects that there must exist certain permanent changes in the net core magnetization (in other words, permanent STT) correspondingly, which is indeed the case as we examine simulations with arbitrary SW propagation. For simplicity, here we show the case under SW excited by ($h_0 = 500$ Oe, $\omega/2\pi = 25$ GHz), see the non-zero $\Delta$ in Fig. 14(c). Thus far, the ever-present excitation of circular mode can be reasonably understood.

As to the elliptical mode (i.e., orbit with periodically changed radius) by SWs with frequencies other than 0.24 GHz, bearing in mind that constant ST torque gives rise to perfect circular motion of core, we reckon that an oscillating ST torque (or periodic STT) is herein demanded. Indeed, oscillations of core component magnetization $m_{z,core}$ can be clearly seen in Fig. 14(c). By FFT analysis, one further finds an oscillating frequency identical to that of the propagating SW and the elliptical motion, see the dominant peak at 25 GHz. The small peak at 0.24 GHz may relate to a periodic distortion of VC due to the large periodic deviation of core away from disk center [81] (note that the motion center of the circular mode is considerably shifted).

To verify the correspondence between circular mode (elliptical mode) and permanent STT (periodic STT), we investigate the driven motion of VC under ac and dc SPC. In Fig. 13(e), we show the recorded trajectory under a small ac SPC of $3 \times 10^{11} \times \sin(5 \times 10^9 \times t)$ $A \cdot m^{-2}$. Note that the core is initially displaced by a field pulse and tend to undergo damped spiral gyration until disk center. Along with the skeletal spiral gyration, small fluctuations at frequencies near 5 GHz are clearly stimulated. Similar to cases in Fig. 10, mixed modes of $5 \pm 0.24$ GHz are observed (see Fig. 13(f)), which in this case can be ascribed to coupling between the intrinsic spiral gyration and the forced synchronous motion by the ac SPC (notably, the absence of peak of 5 GHz may attribute to the overwhelmingly larger amplitude of the 0.24-GHz gyration compared with the 5-GHz motion). Nevertheless, obvious ellipticity effect in each mixed mode can be observed. Similar results are also found in case under SPC containing both ac and dc components, see Fig. 13(g) and 13(h) where we have applied SPC of $(6 \times 10^{10} + 3 \times 10^{11} \times \sin(5 \times 10^9 \times t))$ $A \cdot m^{-2}$. The dc component functions to trim the orbiting radius until new equilibrium, which is in consistence with early studies [77,78,82]. The corresponding $m_{z,core}$ profiles with respect to SPCs $3 \times 10^{11} \times \sin(5 \times 10^9 \times t)$ $A \cdot m^{-2}$ and $(6 \times 10^{10} + 3 \times 10^{11} \times \sin(5 \times 10^9 \times t))$ $A \cdot m^{-2}$ are displayed in Fig. 14(e) and 14(g). Indeed, the ac SPC causes no net changes in $m_{z,core}$ (hence no permanent STT) but oscillations (hence only periodic STT) with frequency same as that of the injecting SPC (see Fig. 14(f)). Thereby we confirm the causality between elliptical mode and periodic STT. On the other hand, the existence of dc SPC independently brings a permanent gain in $m_{z,core}$ (hence permanent STT), and we confirm the causality between circular mode and permanent STT.

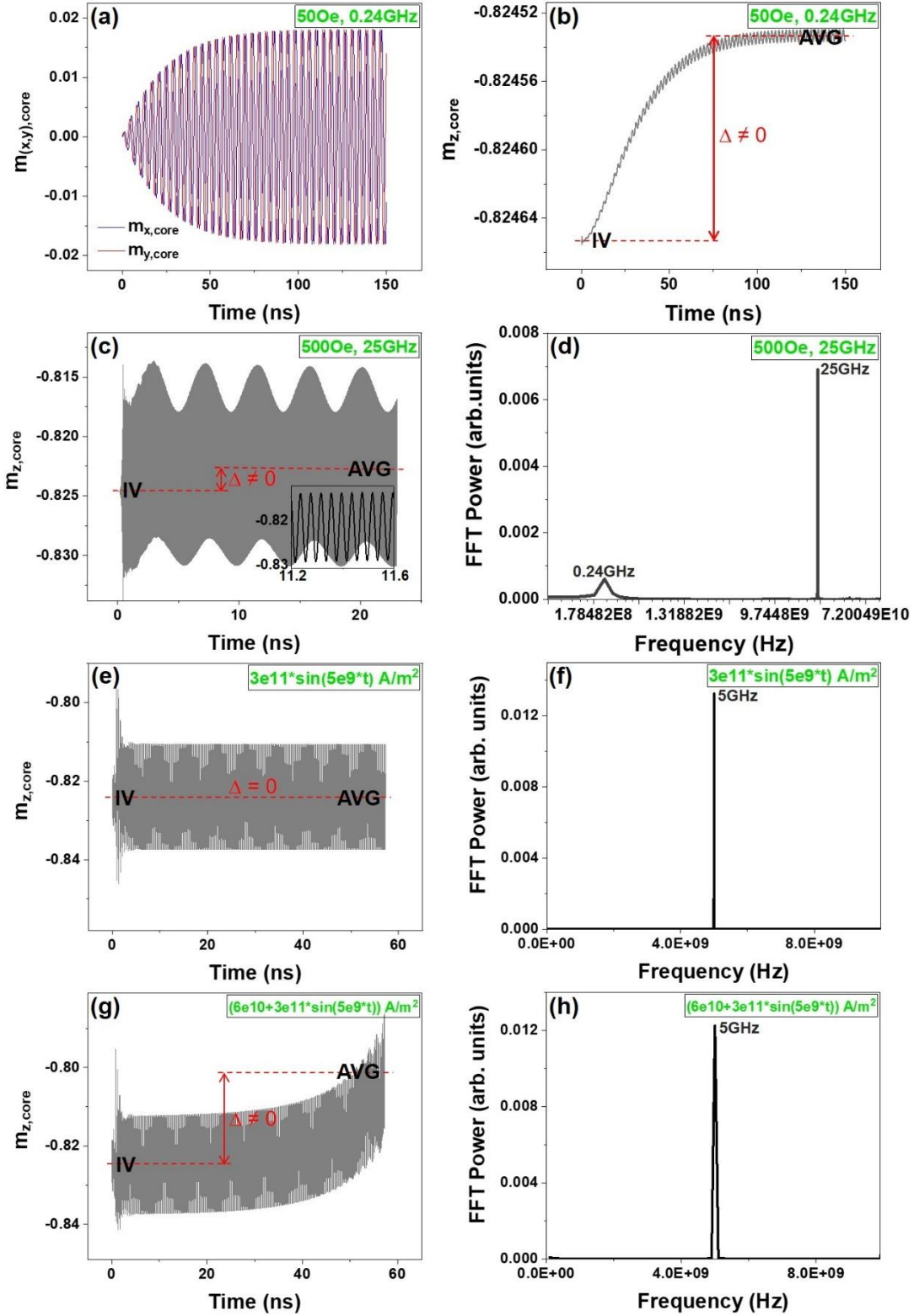

FIG. 14. (a) The x, y components of VC magnetization as a function of time under SW excited based on $h_0 = 50$ Oe, $\omega/2\pi = 0.24$ GHz. The z component of VC magnetization as a function of time under (b) SW excited based on $h_0 = 50$ Oe, $\omega/2\pi = 0.24$ GHz (c) no external stimuli (e) spin polarized current of $3 \times 10^{11} \times \sin(5 \times 10^9 \times t)$ $A \cdot m^{-2}$ (g) spin polarized current of $(6 \times 10^{10} + 3 \times 10^{11} \times \sin(5 \times 10^9 \times t))$ $A \cdot m^{-2}$. IV and AVG respectively abbreviate the initial value and the average of $m_{z,core}$. Δ denotes the difference between them. (d), (f) and (h) are respectively FFT spectrums of (c), (e) and (g).

To acquire comprehensive knowledge about the two modes, we further scan the ($h_0$, $\omega$) space for the essential characters therein. In Fig. 15(a), we show the radius of the circular mode with respect to $h_0$ (or amplitude of SW) at several representative frequencies. The radius initially increases with $h_0$, as expected. However, in each case, there exists certain threshold beyond which the radius starts to drop. This may ascribe to changes in residing potential due to the large displacement of VC from disk center and hence alterant dynamics based on Thiele formulism. Conceivably, a large displacement can originate from three sources: i) large deviation of orbiting center of circular mode due to strong skew scattering and/or back reflection, ii) large amplitude of the elliptical mode, iii) and large amplitude of the circular mode itself. In Fig. 15(c), we demonstrate a case conforming to the first scenario, in which one finds significant reduction of the orbiting radius for a larger deviation. Notably, in addition to the reduced radius, the circular mode may even be deformed and transform into elliptical mode. From Fig. 15(a), one can also sense apparent frequency dependence of the circular mode radius, which has been plotted in detail in Fig. 15(b) for case $h_0 = 300$ Oe. The radius majorly increases in the low band and decreases in relatively higher band with respect to frequency, while accompanied with multiple peaks. Strikingly, in Fig. 15(d) we find exactly coincident peaks in the $\Delta m_{z,core}$ spectrum, which again evidence the permanent STT mechanism of propagating SW in exciting circular modes. The peaks throughout the spectrum, on the other hand, suggests the effect of resonance.

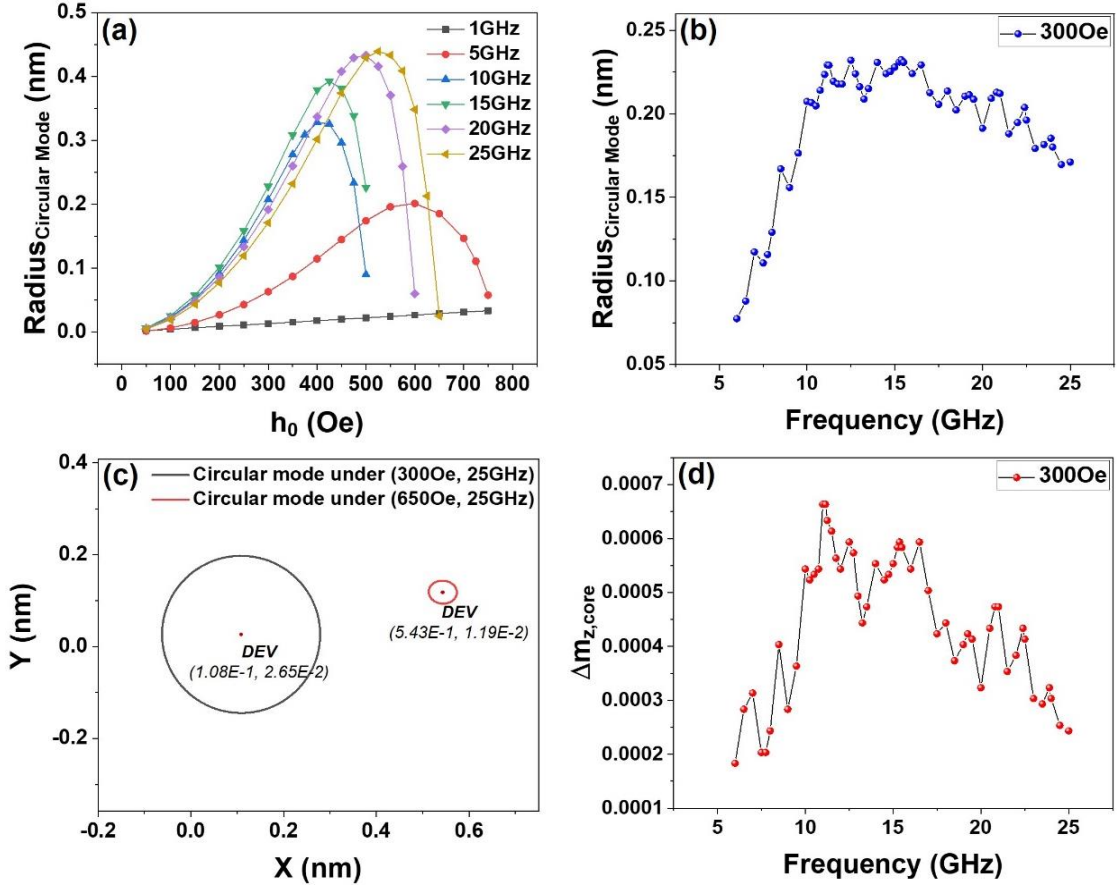

FIG. 15. (a) Radius of circular mode as a function of $h_0$ (before mode collapse) for several representative frequencies. (b) Radius of circular mode as a function of frequency under fixed $h_0$ of 300 Oe. Note that the radius is measured along y axis, and the sampling density is purposely increased around certain $h_0$ and frequencies so as to reveal more details. (c) Comparison of the circular modes excited by ($h_0 = 300\,\text{Oe}$, $\omega/2\pi = 25\,\text{GHz}$) and ($h_0 = 650\,\text{Oe}$, $\omega/2\pi = 25\,\text{GHz}$). (d) Induced $\Delta m_{z,core}$ as a function of frequency under fixed $h_0$ of 300 Oe.

Characters of the elliptical mode (including amplitude, ellipticity and tilting angle) with respect to $h_0$ and $\omega$, are displayed in Fig. 16. Obviously, the amplitude increases with $h_0$ following approximately linear relationship, see Fig. 16(a). The slight bending at large $h_0$ may likewise attribute to the large displacement of VC from disk center. For the same $h_0$, amplitudes at different frequencies further differ, and in Fig. 16(b), one finds dominant peaks at 2, 5, 7, 11, 16 and 21 GHz, suggesting resonant effect of the periodic magnon STT. The ellipticity and tilting angle, on the other hand, remain basically constant as $h_0$ varies, see Fig. 16(c) and 16(e). These two characters also appear to be frequency-dependent. In particular, dominant resonant peaks of 3, 7 and 9 GHz are identified for ellipticity, whereas the tilting angle generally decreases with frequency but with small ups and downs.

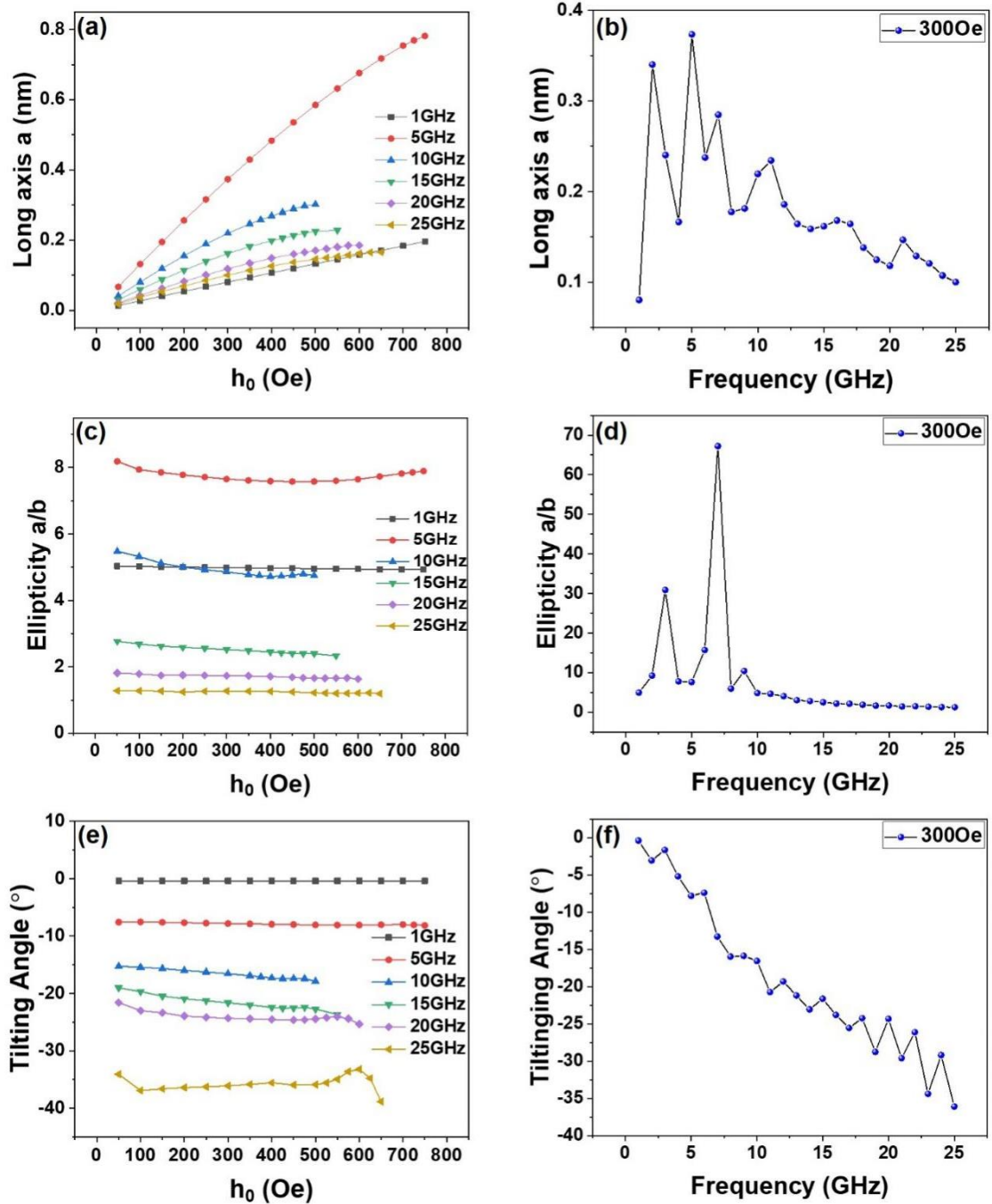

FIG. 16. The long axis of elliptical mode as a function of (a) $h_0$ for several representative frequencies (b) frequency under fixed $h_0$ of 300 Oe. The ellipticity of elliptical mode as a function of (c) $h_0$ for several representative frequencies (d) frequency under fixed $h_0$ of 300 Oe. The tilting angle of elliptical mode as a function of (e) $h_0$ for several representative frequencies (f) frequency under fixed $h_0$ of 300 Oe. (a), (c), (e) are plotted based on the same sampling points as in Fig. 15(a). The sampling points in (b), (d), (f) are evenly spaced with interval of 1 GHz.

## IV. Vortex-based spin wave valve

In view of the multi-scatterings of SW by vortex and limited deviation of VC in translational modes by SW propagation over a wide range of frequencies and intensities

(note that the VC deviation is typically less than 1 nm, see Fig. 15(a)(b) and Fig. 16(a)(b)), vortex is highly promising for realization of SW valve. In Fig. 17, we show a possible design with simple geometry in which VC acts as a switch. The propagating SW can be switched on and off as we shift the position of the core, which can be done using an external dc magnetic field $H_{ext}$ (note that such magnetic field can be easily added by electric current flowing in perpendicularly placed wire). The joint between the disk and the interconnecting nanostrip breaks the tangent magnetization distribution at disk edge, hence VC spontaneously shifts into the north semicircle (note that magnetizations inside strip without strong magneto-crystalline anisotropy align in parallel with the longitudinal direction of the strip so to minimize magnetostatic energy). Thus, in order to block SW, one ought to apply a suitable negative field to pull VC back to disk center. In case of Fig. 17, it is determined to be -5.85 Oe. On the other hand, to channel SW, either large enough positive or negative field (but less than the annihilation field) can be used.

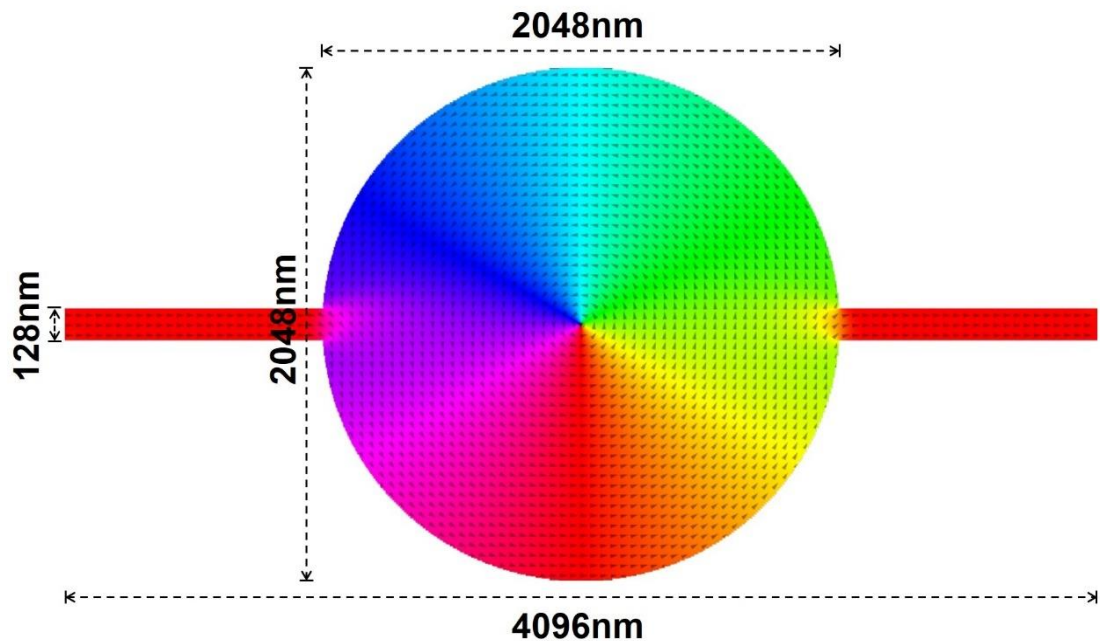

FIG. 17. Schematic of vortex-based SW valve in which propagating SWs are switched on and off by shifting VC position. For simplicity, the diameter of disk is 2048 nm (i.e., R=1024 nm), and the width of interconnecting strip is set to be 128 nm. The switch-off field is determined to be -5.85Oe. In the remaining content, SWs will be excited in the middle of left-side segment of the nanostrip.

The dispersion relation of SW when switching on the valve has been plotted in Fig. 18(b), in which we have largely deviated the core by an external field of 50 Oe, see inset of Fig. 18(b). In contrast to case in absence of disk (see Fig. 18(a)), the higher bands clearly vanish. The lowest band however remains robust and well preserves its parabolic shape, ensuring unchanged properties of the transmitting SWs. It is worth mentioning that we find nearly identical dispersions for different cases respectively under fields of 50 (see Fig. 18(b)), 60, 70, 80, 90 and 100 Oe (see Fig. 18(c)). Considering almost the same textures in the intersection area of disk and nanostrip under 50, 60, 70, 80 and 90 Oe (see Fig. 21(b)) and complete removal of non-colinear texture under an annihilation field of 100 Oe, one can conclude that the dispersion is not markedly affected by fields or textures but other factors. In Fig. 18(d), we investigate the dispersion for a modified geometry in [-R, R] region. As seen, the quadrupled nanostrip width quadruples the bands.

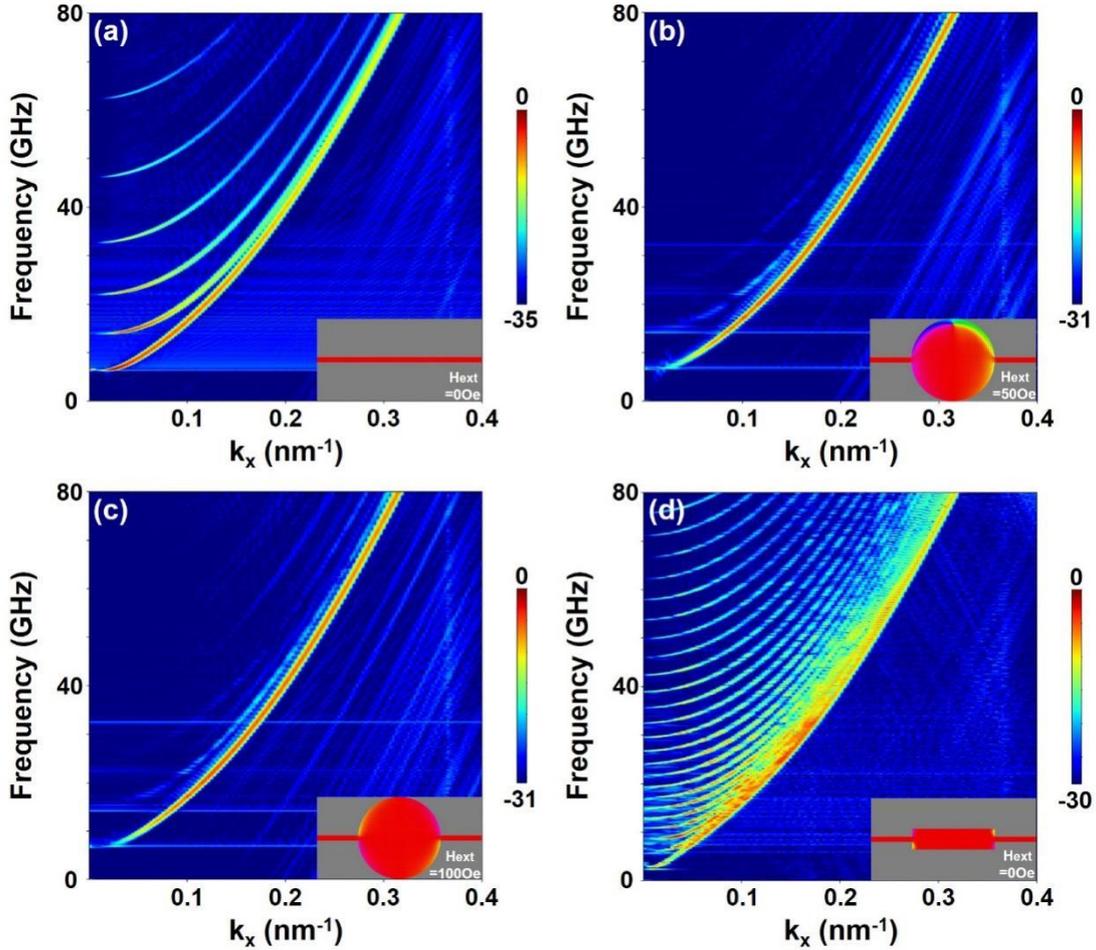

FIG. 18. Dispersion of SW propagation in (a) thin-film nanostrip with width of 128 nm. (b) proposed vortex-based SW valve when switching on with a switch-on field of 50 Oe (c) proposed vortex-based SW valve when the vortex is annihilated under an annihilation field of 100 Oe (d) thin-film nanostrip with quadrupled width in interval [-R, R]. The dispersions are calculated based on interval [-R, R] (R=1024nm) inside the nanostrip.

The spectrum of SW propagation corresponding to Fig. 18(b) is shown in Fig. 19(a) where one observes typical prohibitions of spin waves in low-frequency range of [0, 7] GHz [83,84], fast decay of SWs in mid-frequency range of [7, 17] GHz and most importantly long-distance transmission of SWs in higher-frequency range. Nonetheless, in case when the valve is switched off (i.e., VC locates at the center of disk), VC completely blocks SWs over the entire frequency range, while itself being locally excited, see Fig. 19(b). Based on Fig. 18 and Fig. 19, we confirm the excellent performance of vortex as SW valve.

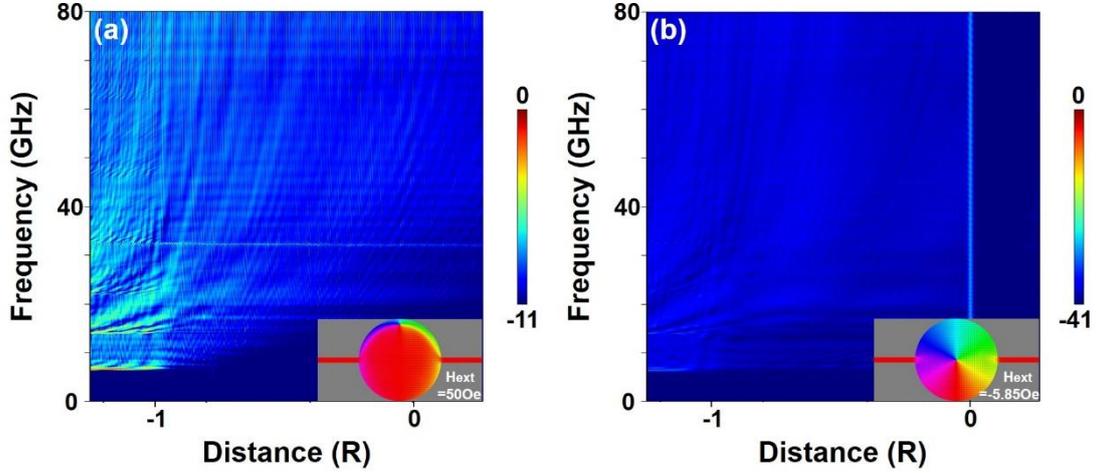

FIG. 19. (a) Spectrum of SW propagation corresponding to Fig. 18(b). (b) Spectrum of SW propagation in proposed vortex-based SW valve when switching off. "0" and "-1" in x axis respectively indicate the center and the left boundary of the disk. The dispersions are calculated based on interval [-R, R] (R=1024nm) inside the nanostrip.

To better meet technological demand in SW computing, it is necessary to examine the wave characters of SW (e.g., the phase of SW) after transmission. In Fig. 20, taking SW excitation under $h_0 = 1\,\text{kOe}$, $\omega/2\pi = 50\,\text{GHz}$ as an example, we show the recorded magnetization as a function of time at locations respectively near left boundary, center and right boundary of the disk (see red curves), in contrast to that of uniformly magnetized nanostrip (see black curves). Note that in both cases, no $H_{ext}$ fields are applied. Clearly, inhomogeneous spin texture causes phase shift of SW. The phase shift has no concern with the direction of non-colinear magnetizations. However, it accumulates even there is no net rotations of in-plane magnetization (i.e., zero winding number), which renovates our early knowledge based on studies of domain wall [32,54,64,85,86]. This further suggests that in order to calculate the exact phase shift of SW after a long-distance propagation, one necessarily follows the track and takes account of every local magnetization inhomogeneity along the trajectory instead of simply evaluates the topological index of winding number, which can be extremely important as it, to a large extent, determines future technological development for the information processing based on SW phase.

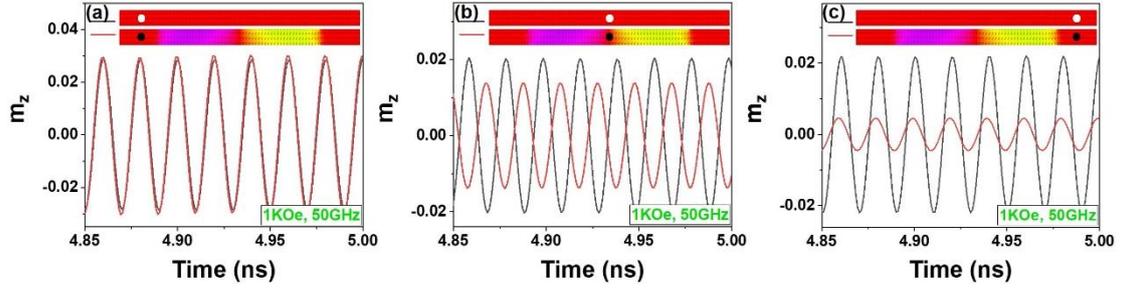

FIG. 20. Comparison between magnetization oscillations (a) at location near left boundary (b) at the center (c) at location near right boundary of disk in valve (see the black dots) and at the same locations (see the white dots) in absence of the disk. SWs are excited based on $h_0 = 1$ kOe, $\omega/2\pi = 50$ GHz. Attached lower strips depict the inhomogeneous magnetizations within the intersection area of disk and interconnecting nanostrip under zero $H_{ext}$, in contrast to the uniform magnetization in upper comparative strips.

Since spin texture can be diversely inhomogeneous in magnetic media, the phase shift with respect to different texture subsequently comes into the next question. In case of vortex-based SW valve, we mostly concern the contouring textures within the intersection area between the disk and the nanostrip under different fields $H_{ext}$ and the corresponding phase shifts of SW (note that different displacements of VC correspond to different spin textures in the intersection area). In Fig. 21(a), we show clear evolution of the spin texture within intersection area subjected to gradually increased $H_{ext}$ along +x direction. A demonstration of different phase shifts under different $H_{ext}$ is given in Fig. 21(c). To be more quantitative, the phase shifts are scanned over a wide range of $H_{ext}$, see Fig. 21(d). Remarkably, in contrast to discrete phase shift reported in many solitonic textures (e.g., domain wall) [54,85], here we obtain smooth phase shift based on continuous modulation of the spin texture, which can result in enhanced flexibility even new functionalities in future device design. Meanwhile, the phase shift clearly covers entire phase range and appears to be asymptotic. The asymptotic line is estimated to be around $8.5\pi$. It is worth mentioning that when $H_{ext}$ reaches the critical annihilation field (in this case, it is determined to be ~96 Oe), the vortex will be saturated (hence a uniform magnetization in intersection area), which causes sharp increase in phase shift. To deepen our understanding regarding the rise of phase shift in inhomogeneous spin texture, in Fig. 21(b), we plotted the $m_y$ profiles corresponding to the textures in Fig. 21(a). Comparison between Fig. 21(b) and Fig. 21(d) suggests that the phase shift closely relates to the dimensional span of non-zero $m_y$ distribution and the amplitude of $m_y$.

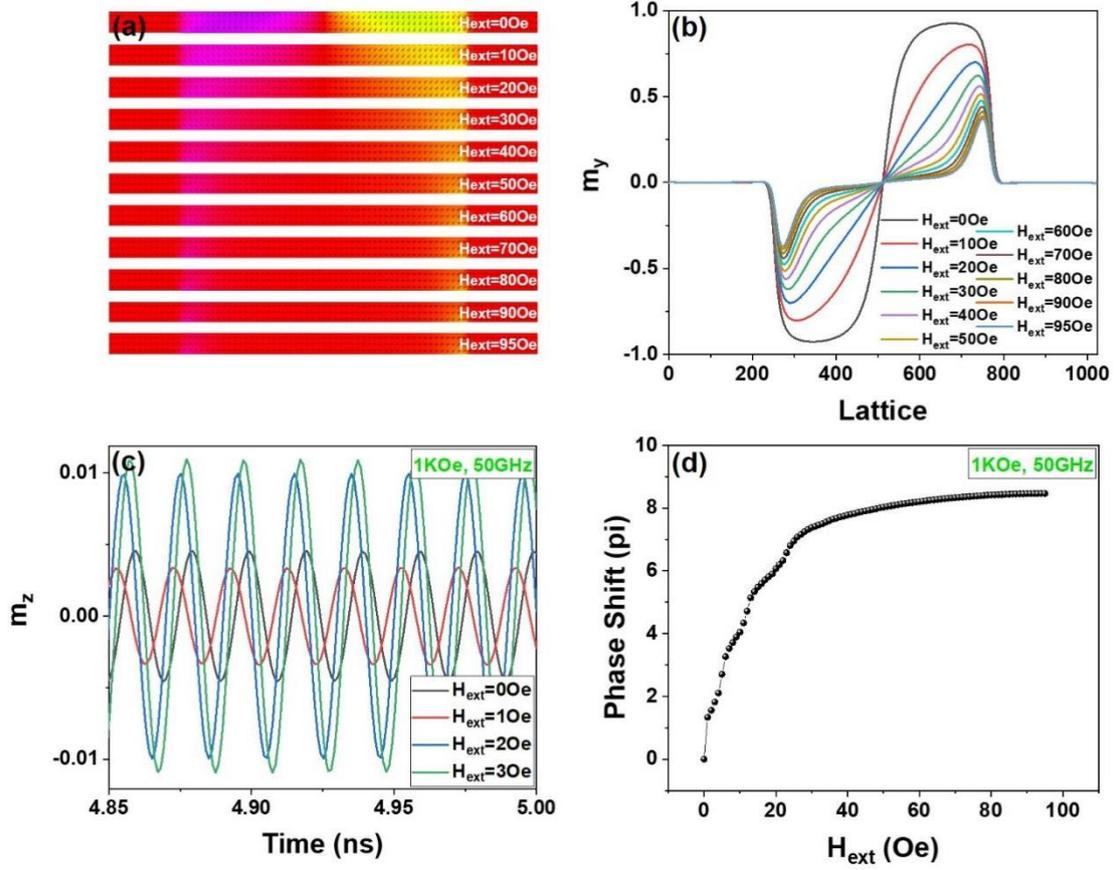

FIG. 21. (a) Inhomogeneous spin textures within the intersection area between the disk and the interconnecting nanostrip in valve under several representative external dc magnetic fields $H_{ext}$. (b) The corresponding $m_y$ profiles with respect to the spin textures in (a). (c) Recorded magnetization oscillations at location same as in Fig. 20(c) for different $H_{ext}$ field of 0, 1, 2, 3 Oe, under SW propagations excited based on $h_0 = 1$ kOe, $\omega/2\pi = 50$ GHz. (d) Phase shift as a function of $H_{ext}$. Note that the phase shift is evaluated with respect to the texture under zero external magnetic field.

To go one step further, in the following, we investigate the partial correlation between phase shift and i). the dimensional span of non-zero $m_y$ ii). amplitude of $m_y$. In Fig. 22(a), vortex-based SW valves with reduced x-axis diameter **b** (or short axis of ellipse) when are simulated under $H_{ext} = 0$ Oe. The $m_y$ profiles in intersection area between the ellipses of different **b** and the interconnecting nanostrip are plotted in Fig. 22(b). As seen, non-zero $m_y$ clearly decreases its x-axis span as **b** decreases (note that zero value of **b** corresponds to uniform magnetization in nanostrip) but maintains nearly unchanged amplitude and shape. The corresponding SWs after transmission and the evaluated phase shifts of the SWs are respectively displayed in Fig. 22(c) and 22(d) from which we confirm the phase shift as a monotone decreasing function of the dimensional span of non-zero $m_y$.

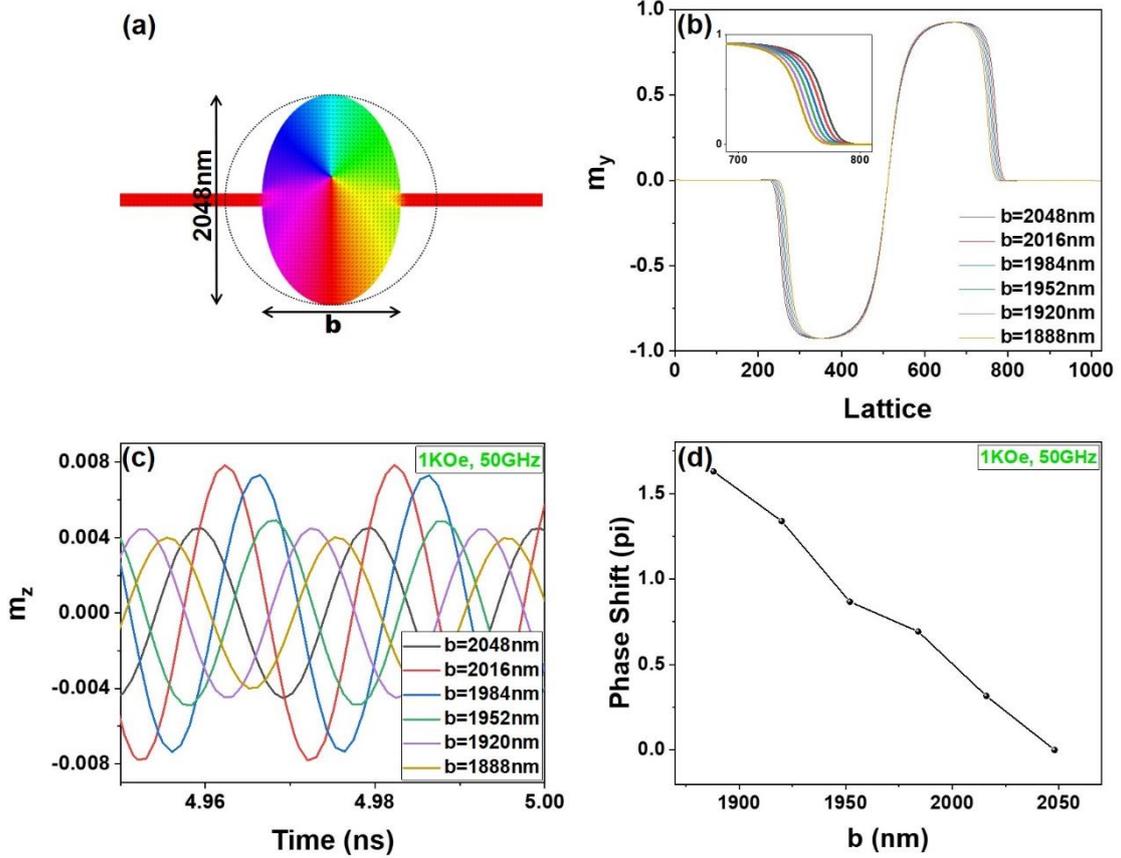

FIG. 22. (a) Schematic of vortex-based SW valve with reduced x-axis diameter **b**. (b) $m_y$ profiles with respect to several representative **b** values. (c) Recorded magnetization oscillations at location same as in Fig. 20(c) under $H_{ext} = 0$ Oe. SWs are excited based on $h_0 = 1$ kOe, $\omega/2\pi = 50$ GHz. (d) Phase shift as a function of **b**. Note that the phase shift is evaluated with respect to the texture under $H_{ext} = 0$ Oe and **b**=2048 nm.

Similar investigation is carried out for vortex-based SW valves with increased y-axis diameter **a** (or long axis of ellipse) under $H_{ext} = 0$ Oe, see Fig. 23. In this case, the amplitude of $m_y$ decreases with increasing **a** (note that an infinite **a** corresponds to uniform magnetization in nanostrip) but maintain approximately unchanged dimensional span of non-zero $m_y$ and the shape of $m_y$ profile. Based on Fig. 23(c) and 23(d), we confirm that phase shift is also a monotone decreasing function of $m_y$ amplitude.

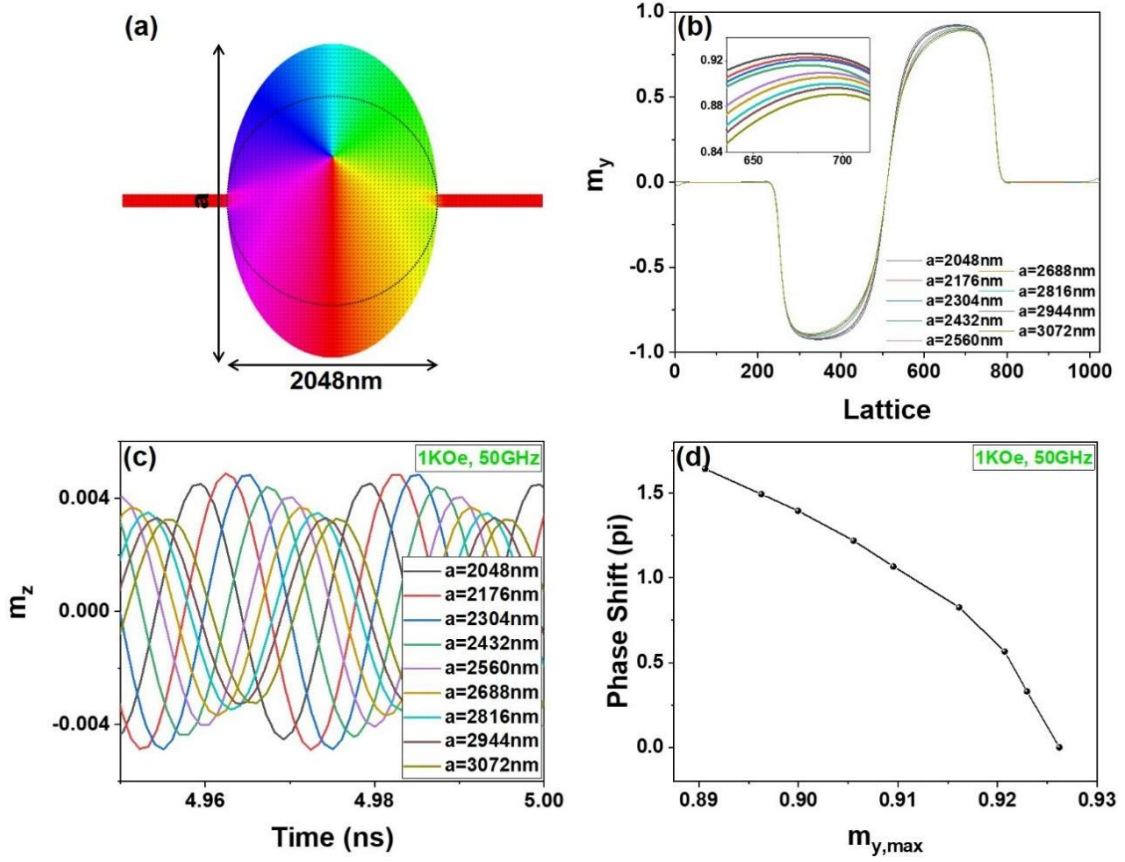

FIG. 23. (a) Schematic of vortex-based SW valve with increased y-axis diameter **a**. (b) $m_y$ profiles with respect to several representative **a** values. (c) Recorded magnetization oscillations at location same as in Fig. 20(c) under $H_{ext} = 0$ Oe. SWs are continuously excited using $h_0 = 1$ kOe, $\omega/2\pi = 50$ GHz. (d) Phase shift as a function of $m_{y,max}$ (i.e., the amplitude of $m_y$). Again, the phase shift is evaluated with respect to the texture under $H_{ext} = 0$ Oe and **a**=2048 nm.

## V.     Conclusion

In conclusion, the interaction between propagating spin wave and Heisenberg vortex in a ferromagnetic disk has been studied. We found three distinct scattering behaviors of spin wave: unilateral skew scattering, symmetric side deflection and back reflection, which respectively associate with the magnetic topology, the local energy density distribution of the texture and the mechanism of linear momentum transfer torque. The skew scattering is well confined within vortex core and is polarity-dependent. The side deflection mainly occurs in out-of-core region and can be tuned through the modification of the size or geometry of the spin system.

On the other hand, the vortex core under steady impinging of magnons exhibits two translational modes: the intrinsic circular mode and a coercive elliptical mode. The former was found to be a result of permanent magnonic spin transfer torque effect, whereas the latter can be ascribed to a process of periodic magnonic spin transfer torque. The essential characters of the two modes generally manifest as a resonant effect in frequency domain. Meanwhile, the amplitude demonstrates strong field-dependence. Coupling of the two modes can further result in frequency comb.

One step further, by virtue of the multiple scatterings of spin wave and immobility of vortex core, we propose a novel spin wave valve based on the vortex, which enables excellent spin wave valving and smooth phase control via inhomogeneity modulation. Noteworthily, the phase shift is determined to be a monotone decreasing function of both amplitude and the dimensional span of inhomogeneity along the contouring texture regardless of its winding number, which renovates previous understanding. Our findings shall deepen the understanding towards the interplay between spin wave and magnetic textures and advance the development of new magnonic devices.

## ACKNOWLEDGMENTS

The authors acknowledge fruitful discussions with Sang-Koog Kim. This work is supported by the National Natural Science Foundation of China (Grants No. 12074332, No. 12074057, No. 11974250, No. 62175155, No. 52172005) and the National Key Research Development Program under Contract No. 2022YFA1402802.